\documentclass[pdflatex,sn-basic]{sn-jnl}

\usepackage{graphicx}%

\usepackage{amsmath,amssymb,amsfonts}%
\usepackage{amsthm}%
\usepackage{mathrsfs}%
\usepackage[title]{appendix}%
\usepackage{xcolor}%
\usepackage{textcomp}%
\usepackage{manyfoot}%
\usepackage{booktabs}%
\usepackage{listings}%

\usepackage{tabularx}
\usepackage{makecell}
\usepackage{multirow}%
\usepackage{threeparttable}
\usepackage{threeparttablex}  
\usepackage{array}
\usepackage{float}
\usepackage{longtable}   

\usepackage{graphicx}
\usepackage{adjustbox}   
\usepackage[linesnumbered]{algorithm2e}
\makeatletter
\AtBeginEnvironment{algorithm}{\setcounter{AlgoLine}{0}}
\makeatletter
\usepackage{breqn}
    
\usepackage{setspace}

\theoremstyle{thmstyleone}%
\theoremstyle{thmstyletwo}%

\theoremstyle{thmstylethree}%

\begin{document}

\title[Article Title]{
Dynamic stacking ensemble learning with investor knowledge representations for stock market index prediction based on multi-source financial data
}


\author[1]{\fnm{Ruize} \sur{Gao}}\email{gaoruize@bimsa.cn}

\author[2]{\fnm{Mei} \sur{Yang}}\email{yangm152@zju.edu.cn}

\author*[3]{\fnm{Yu} \sur{Wang}}\email{yuwang@cqu.edu.cn}

\author[4]{\fnm{Shaoze} \sur{Cui}}\email{shaoze-cui@foxmail.com}

\affil[1]{
    \orgdiv{Digital Economy Lab}, \orgname{Beijing Institute of Mathematical Sciences and Applications}, 
    \city{Beijing}, 
    \country{China}
}

\affil[2]{
    \orgdiv{School of Management}, \orgname{Zhejiang University}, \city{Hangzhou}, \country{China}
}

\affil*[3]{
    \orgdiv{School of Economics and Business Administration}, 
    \orgname{Chongqing University}, 
    \city{Chongqing}, 
    \country{China}
}

\affil[4]{
    \orgdiv{School of Management}, 
    \orgname{Beijing Institute of Technology}, 
    \city{Beijing}, 
    \country{China}
}


\abstract{
The patterns of different financial data sources vary substantially, and accordingly, investors exhibit heterogeneous cognition behavior in information processing.
To capture different patterns, we propose a novel approach called the two-stage dynamic stacking ensemble model based on investor knowledge representations, which aims to effectively extract and integrate the features from multi-source financial data. 
In the first stage, we identify different financial data property from global stock market indices, industrial indices, and financial news based on the perspective of investors. And then, we design appropriate neural network architectures tailored to these properties to generate effective feature representations.
Based on learned feature representations, we design multiple meta-classifiers and dynamically select the optimal one for each time window, enabling the model to effectively capture and learn the distinct patterns that emerge across different temporal periods.
To evaluate the performance of the proposed model, we apply it to predicting the daily movement of Shanghai Securities Composite index, SZSE Component index and Growth Enterprise index in Chinese stock market. The experimental results demonstrate the effectiveness of our model in improving the prediction performance. In terms of accuracy metric, our approach outperforms the best competing models by 1.42\%, 7.94\%, and 7.73\% on the SSEC, SZEC, and GEI indices, respectively.
In addition, we design a trading strategy based on the proposed model. The economic results show that compared to the competing trading strategies, our strategy delivers a superior performance in terms of the accumulated return and Sharpe ratio.
}

\keywords{
Stock market index, multi-source financial data, feature representation, deep learning, dynamic stacking ensemble
}

\maketitle

\section{Introduction}\label{sec1}
Predicting the Stock Market Index (SMI) movement is an important task for both investors and regulators. For investors, accurate prediction is essential for developing effective investment strategies that help to minimize the risk and enhance the return. For regulators, precise SMI prediction enables the crafting of informed policies to maintain the market stability. However, predicting the SMI movement is a difficult and challenging task, since the stock market is inherently non-linear, dynamic, complex, and subject to a high degree of noise \citep{zhongForecastingDailyStock2017,gaoIntegratingSentimentsMultiple2022}.

The data sources influencing a target SMI are diverse. First, historical market data of the target market is an important data source, such as opening price, closing price, highest price, lowest price, trading volume, trading value, and returns \citep{yunPredictionStockPrice2021}. Second, historical data from related markets also play an important role. For example, \citet{hoseinzadeCNNpredCNNbasedStock2019} incorporated multiple SMIs, macroeconomic indicators, crude oil market data, and precious metal prices to predict movements in a target SMI. Their findings demonstrate that integrating numerical data from different markets can significantly improve prediction performance. In addition, industry-level data are also relevant, as the prosperity or downturn of different industries can exert a substantial influence on overall SMI movements. Beyond numerical data, textual information is increasingly utilized, primarily including financial news and social media content. Financial news is typically composed of long, professionally written texts, whereas social media comments are usually short and unstructured contents. Owing to their professionalism, financial news texts tend to attract greater investor attention than the social media comments.

For accurately predicting SMI movements, existing studies have increasingly adopted a multi-source data perspective, proposing advanced methodologies that integrate diverse data sources to enhance the forecasting performance of target SMI \citep{zhongForecastingDailyStock2017,yeShorttermStockPrice2024}.
Among these approaches, the most commonly used can be categorized into three main types: Principal Component Analysis (PCA)-based methods \citep{zhongForecastingDailyStock2017,wengPredictingShorttermStock2018}, tensor-based methods \citep{liRoleSocialSentiment2017,zhangImprovingStockMarket2018,liMultimodalEventDrivenLSTM2021,wangEssentialTensorLearning2023}, and deep learning-based methods \citep{wangCombiningWisdomCrowds2018,xuStockMovementPredictive2020,maMultisourceAggregatedClassification2022,yeShorttermStockPrice2024}. The first type employs techniques such as PCA, fuzzy robust PCA (FRPCA), and kernel-based PCA (KPCA) for dimension reduction. Following this procedure, machine learning methods are applied to learn the relationships between the input variables and stock market prices or indices. The second type uses tensors to represent the information extracted from multimodal data. For example, \citet{liMultimodalEventDrivenLSTM2021} represent three dimensions of the tensor as firm-specific features, professional view features, and sentiment features. Based on the tensor representation, tensor decomposition and reconstruction are used to extract the information. Tensor decomposition is similar to higher-order PCA, while tensor reconstruction can reveal the latent information relationship among various data sources. The third type focuses on designing specialized feature extraction layers tailored to each input, allowing for the effective extraction of relevant features from multimodal data. Subsequently, an integration layer is employed to merge the features extracted from different data modalities, thereby facilitating more accurate stock price prediction. For instance, \citet{maMultisourceAggregatedClassification2022} implement a strategic approach by using a linear layer to extract news sentiment features of the target stock and a graph convolutional layer to capture the news sentiment features of related stocks. This method not only highlights the utility of sophisticated neural network architectures in handling complex data inputs, but also enhances the model's ability to discern and integrate multifaceted market information. 

Although previous studies have adopted a multi-source data perspective and approaches to fuse heterogeneous information and improve prediction performance, several important limitations remain. First, these methods primarily focus on feature-level representation learning and static fusion. As a result, PCA-based techniques often lose essential nonlinear structures during dimensionality reduction, while tensor-based models rely on relatively stable inter-modal relationships that do not align with the time-varying dynamics of real financial markets. Deep learning approaches, despite their stronger feature extraction capability, typically employ fixed fusion architectures that fail to adapt to shifting market conditions. More importantly, most existing models establish a direct nonlinear mapping between inputs and outputs without considering how investors cognitively process multi-source information. In practice, the influence of multi-source financial data on SMI movements is transmitted through investors’ behaviors, i.e., how they identify patterns within each data source, form expectations, and make decisions. Recent studies have also shown that visual information or expert experience (both carrying implicit investor cognition) can significantly enhance forecasting performance \citep{jiangReImaginingPriceTrends2023,caoManVsMachine2024}. Consequently, it is necessary to propose a multi-source financial data fusion framework that incorporates the investor cognition perspective, extracts deep feature representations from various data sources, and dynamically adjusts prediction mechanisms over different time windows.

Therefore, we propose a two-stage dynamic stacking ensemble (TDSE) model based on investor knowledge representations, which aims to effectively extract and integrate the features from multi-source financial data. 
In the first stage, we consider the intrinsic patterns of different data sources by investigating the trend similarity among intraregional SMIs, the industry rotation effect, and the different influence of multiple news providers. To capture the diverse patterns and extract inherent features of global SMIs, industry indices data, and financial news data, we adopt the tailored feature extraction methods, i.e., the multiple branch convolutional neural network, spectral cluster-based multiple branch convolutional neural network, and the recurrent neural network based on evidential rule. Each model is designed to effectively extract deep feature representations from each financial data source. In the second stage, we employ multiple meta-classifiers to effectively learn from the extracted features. Due to the temporal dynamics inherent in the relationship between the extracted features and the SMI movement, we adaptively select the optimal meta-classifier that effectively captures the evolving data patterns in each time period. To verify the effectiveness of the proposed model, we apply it to the prediction of SMI movement in Chinese stock market. The experimental results show that our model outperforms the competing methods in both prediction and economic performance.

The remainder of this paper is organized as follows. Section 2 presents a comprehensive review of the related literature. In Section 3, we delve to the methodology proposed in our study. Section 4 reports the experimental study. In Section 5, we discuss the theoretical and practical implications. Finally, we conclude this paper and present the future work in Section 6.

\section{Related work}

\subsection{Multi-source financial data used in financial prediction}
\subsubsection{Historical market data used in financial prediction}
In financial prediction, the market data plays a pivotal role. It includes various components such as the open price, close price, highest price, lowest price, trading volume, trading value, and returns. These market data reflect the investors' trading behavior. Previous studies have extensively explored the patterns within historical market data and leveraged them to predict the SMI movement \citep{constantinouRegimeSwitchingArtificial2006,orimoloyeComparingEffectivenessDeep2020,yunPredictionStockPrice2021}. The experimental results show that the historical market data has a significant prediction power for the SMI movement. 

With the development of economic globalization, there has been a growing interest in understanding the interaction between different stock markets. Researchers have made efforts to explore the information flow and relationships among multiple SMIs. \citet{marschinskiAnalysingInformationFlow2002} have utilized the transfer entropy method to analyze the interdependencies among international SMIs, revealing valuable insights into the transmission across markets. This exploration has provided the essential theoretical foundation for predicting the target SMI by considering global SMIs. In recent studies, \citet{kiaHybridSupervisedSemisupervised2018} have investigated the prediction power of global SMIs, gold prices, and crude oil data in forecasting the target SMI movement. Their findings emphasize that the inclusion of historical data from other markets significantly improves the prediction performance compared to those solely relying on the historical data of the target stock market. Similarly, \citet{hoseinzadeCNNpredCNNbasedStock2019} have incorporated multiple SMIs, economic data, crude oil market data, and the precious metal prices to predict the target SMI movement. Their results demonstrate that integrating numerical data from different markets leads to improved prediction performance.

In addition to the historical data of the target market and relevant data, the influence of industry data on the movement of the target SMI cannot be overlooked. Within a stock market, industries consist of groups of listed companies that offer similar products or services. Industry indexes, calculated based on the stock prices of listed companies within each industry, reflect the overall development status of the respective industries. Notably, the industry index data exhibit an industry rotation effect, wherein different industries undergo alternating periods of rise and fall \citep{SuZhongGuoAGuShiChangXingYeLunDongXianXiangYanJiu2017}. Previous studies have delved into understanding the impact of this industry rotation effect on the target SMI. For instance, \citet{HeZhongGuoGuShiBanKuaiXianXiangFenXi2001} has discovered a high frequency of industry rotation effect in the Chinese stock market during stable periods and its influence on the target SMI. Furthermore, \citet{PeiZhongGuoGuPiaoShiChangJinRongChuanRanJiQuDaoJiYuXingYeShuJuDeShiZhengYanJiu2019} established a connection between the industry rotation effect and the economic cycle, emphasizing that the effect varies across different time periods. Based on the statistical analysis, \citet{SuZhongGuoAGuShiChangXingYeLunDongXianXiangYanJiu2017} discovered that from 2013 to 2014, the real estate industry had a lead time of 1-2 periods ahead of the steel industry and the household appliance industry. However, the time interval extended to 1-3 periods in 2016.

\subsubsection{Financial news data used in financial prediction}
In the era of rapid information expansion and the vastness of online data, the availability of textual data has experienced an exponential surge. In the finance field, textual data mainly includes financial news and social media data. Financial news texts are crafted by authors equipped with professional financial knowledge and undergo rigorous scrutiny by editors prior to publication. Consequently, due to their inherent professionalism, financial news texts attract greater investor attention compared to the relatively unrestricted nature of social media data. Investors are more inclined to engage with financial news due to its recognized reliability and expert insights.

Given that financial news is a kind of unstructured data, it necessitates transformation into structured data prior to utilization as input for prediction models. One approach involves leveraging word vectors derived from financial news to forecast stock market trends. For instance, \citet{schumakerTextualAnalysisStock2009} utilize techniques such as bag of words, noun phrases, and named entities to represent financial news, illustrating that combining these representations with market data enhances prediction performance. \citet{shynkevichForecastingMovementsHealthcare2016} investigate the impact of word vectors extracted from financial news on stock movements. The experimental results show that the integration of word vectors across multiple industries contributes to improved prediction performance. \citet{namFinancialNewsbasedStock2019} employ word vectors from financial news pertaining to related companies to forecast target stock movements, illustrating that considering the correlation between individual stocks enhances the prediction capabilities of financial news.

While the word vectors can be utilized to as the inputs for predicting the SMI movement, their processing method compromises the interpretability of financial news, posing challenges for investors in making informed decisions solely based on word vectors. Consequently, researchers have turned to sentiment analysis techniques to enhance the interpretability of financial news and facilitate investors' decision-making process. \citet{liEffectNewsPublic2014} emphasize that the financial news sentiment plays a crucial role in the chain connecting news to stock market movement. The experiment results in \citep{liEffectNewsPublic2014} show that utilizing news sentiment for prediction outperforms the use of news vectors at the stock, sector, and stock index levels.

\subsubsection{Fusion of multi-source data in financial prediction}
The related studies on the integration of multi-source data are shown in Table \ref{table:related studies}. We can see from Table \ref{table:related studies} that the numerical data type mainly include market data, technical indicators, fundamental data, macroeconomic data, and exchange rate data. Among these, market data is the most commonly used numerical data, indicating its significance in financial forecasting. On the other hand, when it comes to textual data representation, sentiment analysis is prevalent as the most widely used textual analysis technique. In addition, previous studies often rely on the word embedding method to effectively represent textual data from financial news and social media sources. Moreover, it is noteworthy that researchers typically incorporate multiple data sources, ranging from three to five, in their studies. This demonstrates the recognition of the importance of utilizing multimodal data for improving the prediction performance. 

\begin{table}[!ht]
\centering
\setlength{\belowcaptionskip}{10pt}
\caption{Related studies on the integration of multi-source financial data}
\label{table:related studies}
\footnotesize  
    \begin{tabularx}{\textwidth}{
        >{\raggedright\arraybackslash}X
        p{5cm}
        p{4cm}
        }
    \hline
    Author  & Financial data sources & Fusion methods \\ 
    \hline
    \citet{zhongForecastingDailyStock2017}  & Market data, macroeconomic data, exchange rate data & Feature reduction: PCA model  \\
    
    \citet{liRoleSocialSentiment2017} & Market data, Financial news: Noun term vector \& sentiment; Social media data: sentiment & Tensor decomposition \& reconstruction \\
    
    \citet{wengPredictingShorttermStock2018} & Market data, technical indicators, Google trends, Wikipedia hits, Financial news: sentiment & Feature reduction: PCA model \\ 
    
    \citet{zhangImprovingStockMarket2018} & Market data, Financial news: event; Social media data: sentiment & Coupled matrix and tensor decomposition \\
    
    \citet{wangCombiningWisdomCrowds2018} & Technical indicators, Social media data: sentiment & Deep random ensemble learning \\ 
    
    \citet{xuStockMovementPredictive2020} & Market data, Social media data: Word embedding & Deep learning-based feature extraction \\ 
    
    \citet{liMultimodalEventDrivenLSTM2021} & Market data, fundamental data, Financial news: sentiment  & Tensor decomposition and reconstruction \\ 
    
    \citet{maMultisourceAggregatedClassification2022} & Market data, technical indicator, Financial news: Word embedding & Deep learning-based feature extraction \\ 
    
    \citet{wangEssentialTensorLearning2023} & Fundamental data, company business conditions, Financial news: sentiment;  & Feature reduction: Tensor robust PCA \\ 
    
    \citet{liResidualLongShortterm2023} & Volume and price, technical, and macro data & Feature fusion residual LSTM \\ 
    \citet{gaoInvestorSentimentStock2024} & Market data, Social media data: investor sentiment & OLS \\
    \citet{dongDistillingWisdomCrowds2024} & Market data, Online comments: word embedding & BERT-BiLSTM model \\ 
    \citet{zhaoStockReturnForecasting2025} & Market data, macroeconomic data, other types of market data (Bond, Foreign exchange, Commodity) & Random forests\\
    \citet{shenForecastingTourismStock2026} & Market data, Macro data, Social media & Deep learning\\
    \hline
    \end{tabularx}
\end{table}

When it comes to the integration process of multi-source financial data, it can be broadly categorized into two types: feature reduction methods and feature extraction methods. Among the feature reduction methods, the PCA method effectively reduces dimensions from multi-dimensional vector data. On the other hand, the tensor-based method represents the input data using tensors and utilizes decomposition and reconstruction techniques to capture the interaction of multi-source data. As for the feature extraction method, the deep learning method is an effective method to integrate the multi-source data. Based on the end-to-end mechanism, the deep learning method gradually extracts features from diverse data sources and utilizes them to predict the target SMI movement.

\subsection{Prediction techniques using multi-source financial data}
Based on previous review, we find that various financial prediction methods based on either numerical data or textual data commonly utilize the SVM model, the ANN model, and other individual techniques, which are suitable for addressing unimodal data. However, when it comes to handling the complexities within multimodal financial data, the hybrid model has become a widely adopted approach in financial prediction. These hybrid models consist of various techniques such as the PCA-based machine learning model, the tensor-based machine learning model, and advanced deep learning models. By combining these diverse methods, the hybrid model efficiently captures the interaction between different data modalities for improving the prediction performance.

The first type of model, known as the PCA-based machine learning model, combines the PCA method and machine learning techniques such as SVM and ANN model to forecast the SMI movement. The PCA method is a kind of feature reduction method, essentially mapping the original feature space of multimodal data into a new feature space by constructing a new coordinate system. The transformation enables the generation of new features that effectively reduce redundant information. For example, \citet{zhongForecastingDailyStock2017} employ the PCA, the KPCA, and the FRPCA model to reduce dimensions of market data, macroeconomic data, and exchange rate data. The experimental results show that combining PCA-type method with ANN models yields superior prediction performance for forecasting the S\&P 500 index movement compared to other models. Similarly, \citet{wengPredictingShorttermStock2018} utilize the PCA model to extract features from diverse sources such as market data, technical indicators, financial news, Google trends, and Wikipedia hits. Subsequently, they employ the Boosting Regression Tree (BRT) model to forecast the movement of multiple stocks. Their results showcase the efficacy of the proposed model, outperforming other competing methods in terms of prediction accuracy. These studies investigate the potential of PCA-based approaches in extracting informative features from multimodal financial data.

The PCA method primarily operates on vector-based data representations. In contrast, the second type of model, known as the tensor-based machine learning model, utilizes the tensor as a fundamental data structure \citep{leeHierarchicalMultimodalFusion2025}. Tensors extend the notion of vectors and matrices by incorporating higher-order dimensions. Tensor-based methods have gained attention in the field of multimodal financial data analysis due to their ability to capture intricate interactions among different modalities. For example, \citet{liRoleSocialSentiment2017} employ the tensor-based model to integrate market data, financial news, and discussion boards. Their results show that the proposed tensor-based model outperforms the Tucker model the iterative optimization model, demonstrating the effectiveness of tensors in capturing the complex relationships among different modalities. \citet{zhangImprovingStockMarket2018} utilize a tensor-based decomposition and reconstruction method to combine stock quantitative data, social media data and the web news data. Their results show that the proposed model outperforms traditional approaches such as SVM and PCA+SVM model in terms of accuracy. \citet{wangCombiningWisdomCrowds2018} devise a novel approach that leverages tensor-based representations for fundamental data, sentiment features, and market information. They further propose an attention-based long short-term memory model to forecast stock movements. The experimental results show that their proposed model is superior to competing methods in terms of accuracy the Matthews correlation coefficient. Additionally, \citet{liMultimodalEventDrivenLSTM2021} employ a tensor transformation method to represent market data and news data. They subsequently employ the LSTM model to forecast the stock movement. The experimental results show that their proposed model outperforms other competing models in terms of classification performance and simulated investment returns. \citet{snaselGeneralizationMultisourceFusionbased2024} proposed a novel framework based on multi-source data fusion and decision-level fusion for stock selection problems. The framework used vector-based data representation and adopted dynamic time warping (DTW) and a customized loss function to improve the accuracy of stock time series prediction. Collectively, these studies investigate the potential of tensor-based models in effectively integrating multimodal financial data and improving prediction accuracy across various domains within the financial field.

The third type of model, namely the advanced deep learning model, emulates the cognitive processes of human behavior. These models begin by extracting low-level and simple features from raw data and gradually progresses to higher-level and abstract features by increasing the number of layers. Deep learning models have been widely used to integrate multimodal financial data in related studies. For example, \citet{xuStockMovementPredictive2020} propose an attention mechanism-based deep learning approach to integrate market data and social media data. Their experimental results demonstrate that the proposed model outperforms competing methods, showcasing its superior prediction performance. Similarly, \citet{wangCombiningWisdomCrowds2018} introduce a deep random subspace ensemble method for fusing technical indicators and social media data. The results verify the effectiveness of their proposed model in effectively combining these two types of data. In another study, \citet{maMultisourceAggregatedClassification2022} employ a comprehensive approach by extracting quantitative indicators of the target stock, news features of the target stock, and news features of important related stocks. They subsequently utilize a Bidirectional Long Short-Term Memory (BiLSTM) model to forecast stock price movements. Their experimental results show that the proposed model outperforms other competing methods in terms of classification performance and financial evaluation metrics. \citet{liResidualLongShortterm2023} propose a model based on feature fusion residual long short-term memory network (FFRL) for predicting the Chinese stock market by combining multi-source and multi-frequency information such as volume and price, technical, and macro data. Experimental results show that FFRL has significant performance improvement over traditional deep learning models. In addition, \citet{dongDistillingWisdomCrowds2024} develop a BERT-BiLSTM prediction model based on multi-source data fusion and word embedding technology to convert user comments in online communities into feature vectors. The model performed well in the price prediction of the constituent stocks of the SSE 50 Index, significantly outperforming traditional methods. Based on the above studies, we find that the utilization of deep learning techniques provides a robust framework for integrating multimodal data and improving the prediction performance.

\subsection{Our contributions}
This study contributes to the field from the following aspects.

(1) We adopt an investor-centered perspective to comprehensively account for the intrinsic patterns underlying various influencing factors and develop an innovative feature extraction framework to represent heterogeneous data sources. 
The learned feature representations allow the model to uncover and exploit the distinctive patterns within each data source, enabling the extraction of more informative and prediction-relevant features for forecasting the target SMI movement.

(2) We propose a novel dynamic stacking ensemble model that effectively integrates the features extracted from multi-source financial data. By considering the temporal dynamics inherent in the relationship between the extracted features and the target SMI movement, our model dynamically selects the optimal meta-classifier for each time period. 

(3) We develop a stage-by-stage optimization method for optimizing the proposed model and enhancing its generalization ability. The staged optimization process allows us to fine-tune the model hyper-parameters in a systematic manner. 
The staged optimization approach helps to reduce the complexity of the solution space.

\section{Methodology}
\subsection{Problem statement}
There are various factors that exert a significant influence on the SMI movement. Among these data sources, three stand out as particularly significant: global SMIs data $\boldsymbol{X}_{1, t}=\{x_{11, t}, x_{12, t}, \ldots, x_{1 n_1, t}\}$, industry index data $\boldsymbol{X}_{2, t}=\{x_{21, t}, x_{22, t}, \ldots, x_{2 n_2, t}\}$, and financial news data $\boldsymbol{X}_{3, t}=\{x_{31, t}, x_{32, t}, \ldots, x_{3 n_3, t}\}$. It is noteworthy that these data sources exhibit different modalities: global SMIs data and industry index data are numerical, whereas financial news data is textual. When investors make decisions based on these data, they assess future market performance by analyzing overall patterns within specific data categories. Therefore, the ability to recognize and leverage the intrinsic patterns from each data source is crucial in enhancing the prediction performance of the target SMI movement. However, discovering overall patterns from the perspective of investors, capturing these distinctive features, and integrating multimodal data within a prediction model present significant challenges. Overcoming these challenges is essential for improving the accuracy and effectiveness of the predictive model.

To address the above challenges, we propose a two-stage dynamic stacking ensemble model consisting of a feature extraction model and a SMI prediction model. Specifically, we observe a strong intraregional correlation within global SMIs data, an industry rotation effect within the industry index data, and influence differentiations from various news providers. Effectively incorporating the inherent intrinsic patterns of these data sources and extracting their features for accurately forecasting the target SMI movement has become a pivotal challenge. To address this, we employ three distinct feature extraction processes, as outlined by the following equations: $fe_{1, t}=F E_1(\boldsymbol{X}_{1,t})$, $fe_{2,t}=F E_2(\boldsymbol{X}_{2,t})$, $fe_{3, t}=FE_3(\boldsymbol{X}_{3,t})$. 
The $FE_1$ is the multi-branch convolutional neural network (MBCNN) model, where the convolutional units within each branch effectively extract features from intraregional SMIs. By the convolutional operation, which is based on parameter sharing, these units facilitate the integration of regional SMI similarities into the feature extraction process. 
The $FE_2$ is the spectral clustering based MBCNN (SC-MBCNN) model, where the spectral cluster algorithm is first used to form multiple industry clusters based on the industry index data. Based on these established industry clusters, the subsequent utilization of the MBCNN model enables the extraction of comprehensive features that can capture the industry rotation effects. 
The $FE_3$ is the recurrent neural network based on evidential reasoning rule (RNN-ER), where the RNN model is first used to construct base classifier for each news provider. The approach effectively captures the varying prediction capabilities of different news providers. Subsequently, the ER rule is used to integrate the extracted features from multiple news providers, enhancing the overall feature representation. These feature extraction processes serve as crucial steps in capturing the underlying patterns of each data source, facilitating the subsequent prediction to make accurate forecasts. 

The extracted features are then effectively combined through the utilization of the prediction model, as depicted by the following equation: 
\begin{equation}
    y_{t+1}=f(fe_{1, t}, fe_{2, t}, fe_{3, t}),
\end{equation}
where the $f$ is the dynamic stacking ensemble function which can capture the temporal dynamics inherent in the relationship between the extracted features and the target SMI movement. The $y_{t+1}$ is the target SMI movement, which is calculated by the following equation:
\begin{equation}
    y_{t+1}=\frac{Close_{t+1}-Close_t}{Close_t},
\end{equation}
where $Close_t$ represents the close price on the $t\text{-}th$ trading day.

\subsection{Model framework}
\begin{figure}
    \centering
    \includegraphics[width=1\linewidth]{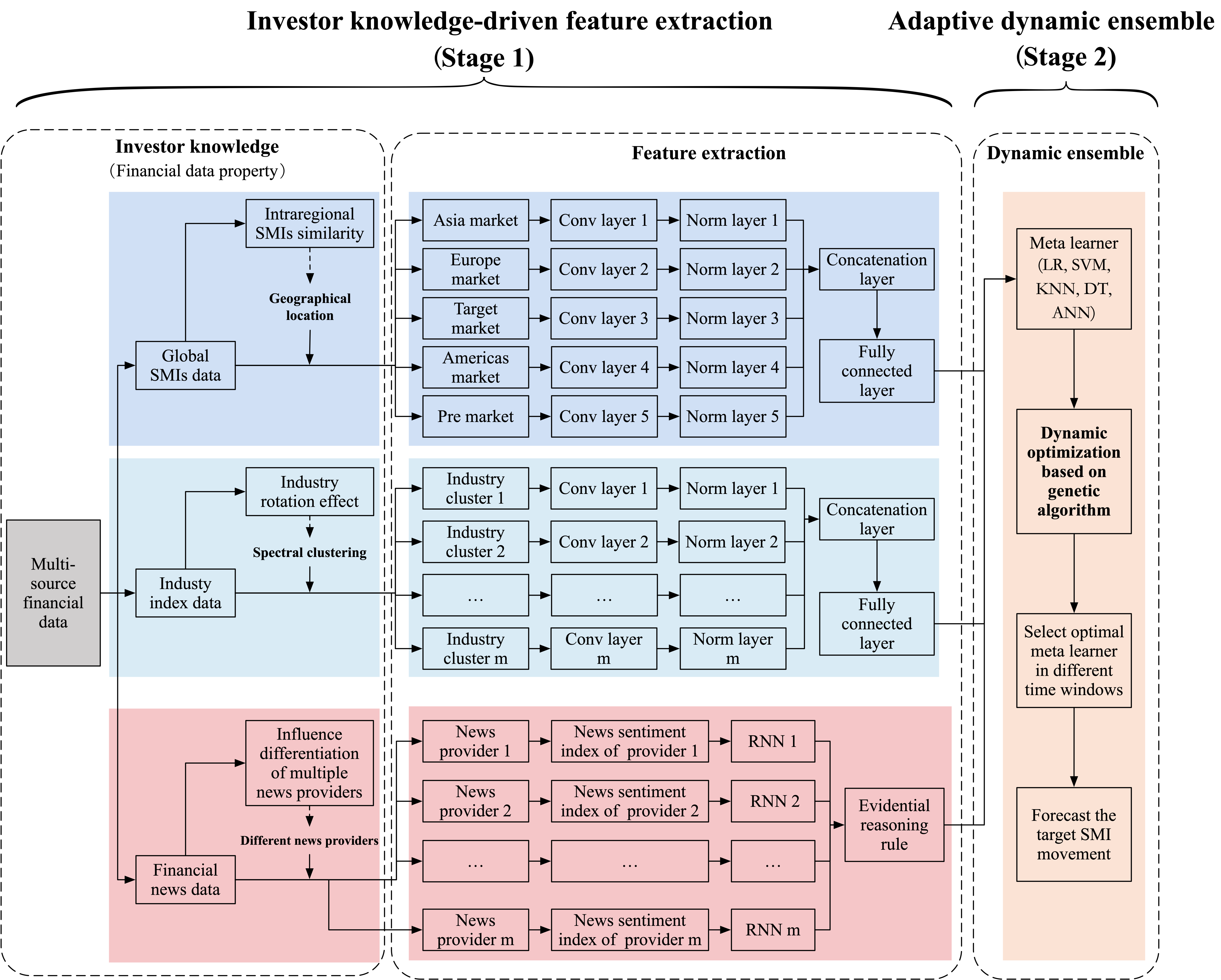}
    \caption{Framework of the proposed model}
    \label{fig:TDSE-DL}
\end{figure}

As previously discussed, forecasting the target SMI movement using multimodal financial data presents a significant challenge due to the distinctive inherent patterns within each data source. To consider these intrinsic patterns inherent in each data source and effectively extract informative features from diverse data sources, we propose the TDSE model for integrating three essential data sources: global SMIs data, industry index data, and the financial news text. The framework of our proposed model is shown in the Fig. \ref{fig:TDSE-DL} which comprises two stages: investor knowledge-driven feature extraction (Stage 1) and adaptive feature fusion (Stage 2). 

In Stage 1, we first analyze the property of different financial data sources from the investors' perspective, and then we adopt three suitable model to extract features from them. Specifically, the MBCNN model is applied to capture features related to intraregional SMIs similarity, allowing us to uncover correlations within the global financial landscape. The SC-MBCNN model focuses on extracting features of industry rotation effects, providing insights into the cyclic nature of specific sectors. The RNN-ER model is utilized to extract features associated with the influence differentiation among multiple news providers, enabling a comprehensive understanding of the impact of diverse news sources. The systematic approach ensures that the feature extraction process utilizes investor knowledge to uncover the intrinsic patterns of each data source, thereby providing valuable insights for accurately predicting the target SMI movement.

In Stage 2, the features extracted in Stage 1 are adaptively utilized in a dynamic stacking ensemble model for feature fusion and target SMI movement prediction. Specifically, we employ five different meta-classifiers: LR, SVM, KNN, DT, and ANN. The rationale behind selecting multiple meta-classifiers stems from their diverse strengths and suitability across different time periods. As each time period necessitates the most appropriate meta-classifier, the selection of meta-classifiers has an important impact on the prediction performance. Based on the proposed model, we effectively aggregate the extracted features, ensuring their synergistic contribution to enhancing the overall prediction performance.

\subsection{Stage 1: Investor knowledge-driven feature extraction}
In this section, we describe the process of investor knowledge-driven feature extraction (Stage 1), which primarily involves the feature extraction of global SMIs, industry data, and financial news. 

\subsubsection{Feature extraction of the global SMIs}
The first data source utilized in our study is the global SMIs data. Within this data, we observe a notable similarity among intraregional SMIs, i.e., the SMIs in a region tend to have the similar movement. Taking into consideration the unique properties of this data source is not only crucial but also highly beneficial in uncovering essential information that can significantly enhance the prediction performance. Therefore, to effectively extract the similarity features of intraregional SMIs, we employ the multiple branch convolutional neural network in this study, which is described as follows.  

Based on the geographical location, we divide the market into five branches for the feature extraction process: Asia Market, Europe Market, Americas Market, Target Market, Pre Market. For the SMIs in each region market, we use some of $R_{O\text{-}C}$, $R_{C\text{-}C}$, and $R_{C\text{-}O}$ as inputs of the model. The $R_{O\text{-}C}$, $R_{C\text{-}C}$ and $R_{C\text{-}O}$ respectively represents the returns of $Open\text{-}to\text{-}Close$, $Close\text{-}to\text{-}Close$ and $Close\text{-}to\text{-}Open$. The detailed explanation of the used SMIs in this study is illustrated in the Appendix A.

For the $i\text{-}th$ branch, the input vector is represented as $\boldsymbol{x}^{(i)}=\{x_1^{(i)}, x_2^{(i)}, \ldots, x_{n_i}^{(i)}\}$, where $x_{n_i}^{(i)}$ is the $n_i\text{-}th$ SMI of the $i\text{-}th$ branch. Given the convolutional kernel $\boldsymbol{w}^{(i)}=[w_1^{(i)}, w_2^{(i)}, \ldots, w_F^{(i)}]$ where $F$ represents the length of convolutional kernel, the convolutional operation is used to extract the similarity feature of intraregional SMIs. The reason why selecting the convolutional operation is due to the characteristic of local relevance and wight sharing, which can help capturing the similarity of intraregional SMIs. The calculation process of convolutional operators is shown in the following equation:
\begin{equation}
    V_j^{(i, l)}=\theta(\sum_{k=1}^F w_k^{(i)} x_{j+k-1}^{(i)}),
    \label{eq:convolutional operators}
\end{equation}
where the $V_j^{(i,l)}$ is the value of the $j\text{-}th$ column of the $l\text{-}th$ layer for the $i\text{-}th$ branch, $w_k^{(i)}$ is the weight of the $k\text{-}th$ column of convolutional kernel, $\theta$ is the Rectified Linear Unit (ReLU) activation function $\theta(x)=\max (x, 0)$. 

To enhance the efficiency of model training while ensure the convergence of the model towards optimal performance, a normalization layer is employed to effectively normalize the distribution of the feature extraction layer. Given the obtained vector of the $l\text{-}th$ layer $\boldsymbol{V}^{(i, l)}=\{V_1^{(i, l)}, V_2^{(i, l)}, \ldots, V_k^{(i, l)}, \ldots, V_E^{(i, l)}\}$,  the value of each dimension is normalized by the following equation: 
\begin{equation}
    \hat{V}_k^{(i, l+1)}=\frac{V_k^{(i, l)}-\mu_V}{\sigma_V},
\end{equation}
where $\mu_V$ is the mean value, $\sigma_V$ is the standard deviation, $\hat{V}_k^{(i, l+1)}$ is the value of the $k\text{-}th$ dimension of the $l+1$ layer. Since the meaning of the feature expression of this layer is likely to be changed during the normalization process, in order to ensure that the transformation of the feature achieves an identity change, the following equation is used to solve the problem:
\begin{equation}
    Z_k^{(i, l+1)}=\gamma_k^{(i)} \hat{V}_k^{(i, l+1)}+\beta_k^{(i)},
\end{equation}
where $Z_k^{(i, l+1)}$ represents the normalized value of the $k\text{-}th$ dimension in the $l+1$ layer, the $\gamma_k^{(i)}$ and the $\beta_k^{(i)}$ are the model parameters to be trained.

Based on the feature extraction of the convolutional layer and the normalization layer, the output of the $l+1$ layer in the $i \text{-} th$ branch is $\boldsymbol{Z}^{(i, l+1)}=[Z_1^{(i, l+1)}, Z_2^{(i, l+1)}, \ldots, Z_k^{(i, l+1)}, \allowbreak \ldots, Z_E^{(i, l+1)}]$. Given $m(m=5)$ branches, there exists new features in the $l+1$ layer $\boldsymbol{Z}^{(1, l+1)}, \boldsymbol{Z}^{(2, l+1)}, \ldots, \boldsymbol{Z}^{(m, l+1)}$. Based on the generated new features of multiple branches, we employ the concatenation layer to fuse these features, which forms the input features of the $l+2$ layer $\boldsymbol{V}^{(l+2)}=[\boldsymbol{Z}^{(1, l+1)}, \boldsymbol{Z}^{(2, l+1)}, \ldots, \boldsymbol{Z}^{(m, l+1)}]=[V_1^{(l+2)}, V_2^{(l+2)}, \ldots, V_n^{(l+2)}]$. The fully connected layer is connected to the concatenation layer. The reason why we employ the fully connected layer here is due to its ability to extract global features from the multiple branches. Given the input vector $V^{(l+2)}$ and the ReLU activation function $\theta$ for the fully connected layer (the $l+3$ layer), the output of the $i \text{-} th$ neuron of the $l+3$ layer is calculated by the following equation:
\begin{equation}
    V_i^{l+3}=\theta(\sum_k w_{k, i}^{l+2} V_k^{l+2}),
\end{equation}
where $w_{k, i}^{l+2}$ is the connected weight between the $k\text{-}th$ neuron of the $l+2$ layer and the $i\text{-}th$ neuron of the $l+3$ layer. Finally, the extracted features are obtained by the output layer with the softmax activation function. There are $n(n=2)$ types where the prediction probability of the $i\text{-}th$ type type is calculated by the following equation: 
\begin{equation}
    p_i=\frac{e^{z_i}}{\sum_{j=1}^n e^{z_j}},
    \label{eq:output of MBCNN}
\end{equation}
where the $z_j$ represents the output value of the $j\text{-}th$ neuron in the output layer.

Based on the feature extraction process from the global SMIs data, the prediction probability results $\boldsymbol{p}_G$ are determined by the MBCNN model, which can be further utilized to further improve the prediction performance in the Stage 2.

\subsubsection{Feature extraction of the industry data}
In this study, the industry index data is the second valuable data source. Based on the previous studies \citep{SuZhongGuoAGuShiChangXingYeLunDongXianXiangYanJiu2017,HeZhongGuoGuShiBanKuaiXianXiangFenXi2001}, there exists industry rotation effect for the industry index data, which refers to the stock market pattern where different industries have risen or fall in turn for a period of time. The industry rotation effect can provide important information for the target SMI prediction. For the industry rotation effect of the industry index data, we employ the SC-MBCNN model to extract the feature of industry rotation effect. 

To effectively extract the feature of industry index data, we design two blocks for the feature extraction process: industry cluster division based on spectral cluster algorithm (Block 1) and feature extraction of industry rotation effect (Block 2). The steps of feature extraction process are described as follows. 
First, based on the industry index and target SMI data, we divide the dataset into multiple time periods, thereby forming industry index matrix for each period, i.e., $\boldsymbol{II}_{n \times x}$ where the $n$ is the number of industry index and the $l$ is the length of selected time periods. And then, we employ the spectral clustering (SC) algorithm to form industry clusters based on the industry index data within each time window. The SC algorithm is proposed by \cite{ngSpectralClusteringAnalysis2002}, which regards the samples as points in the space and connects the sample points by edges. The edge weights between the farther sample points are assigned lower while the edge weights between the closer sample points are assigned higher. And then, the sample points are divided into subgraphs by the graph partitioning method, where each subgraph is a cluster. Finally, the sum of edge weights in different subgraphs is as high as possible, and the sum of edge weights between subgraphs is as low as possible. Because the SC algorithm has the advantages of strong adaptability to data distribution and small amount of calculation, it has been successfully applied to the financial time series data \citep{liUndirectedDirectedNetwork2022}.

We describe spectral clustering based the industry cluster division process as follows. Given the industry index matrix $\boldsymbol{II}=\{F_1, F_2, \ldots, F_n\}$ where $F_i$ is the attribute of the $i\text{-}th$ industry, i.e., $F_i=\{\text{Value}_{i,1}, \text{Value}_{i,2}, \ldots, \text{Value}_{i,l}\}$ where $l$ is the length of time periods. The industry index features are connected to form an undirected weighted graph $G(\boldsymbol{II}, A)$ where $A_{ij}$ is the similarity between the $i\text{-}th$ industry index feature and the $j\text{-}th$ industry index feature, which is calculated by the following equation: 
\begin{equation}
    A_{i j}=\exp (-\frac{d\left(F_i, F_j\right)}{2 \sigma^2}),
    \label{eq:similarity between industries}
\end{equation}
where $exp()$ is the exponential function with base of natural constant, $\sigma$ determines the width of neighborhoods, the $d(F_i, F_j)$ is the Euclidean distance between two industry index features, which is calculated in the following equation: 
\begin{equation}
    d(F_i, F_j)=\sqrt{\sum_{k=1}^l(\text { Value }_{i, k}-\text { Value }_{j, k})^2}.
    \label{eq:Euclidean distance}
\end{equation}
Based on the Eq. (\ref{eq:similarity between industries}) and (\ref{eq:Euclidean distance}), when the scale parameter $\sigma$ remains constant, the closer the distance between two industry indexes, the greater the connection weights. Based on the constructed undirected weighted graph, the method of graph partitioning is used to maximize the sum of weights within each subgraph and minimize the sum of weights between subgraphs. The details of the SC algorithm are shown in the Algorithm \ref{algo:spectral algorithm}. 

\begin{algorithm}[H]
    \SetKwInOut{Input}{Input}
    \SetKwInOut{Output}{Output}
    \caption{Spectral cluster algorithm used for the industry index matrix}
    \label{algo:spectral algorithm}
    \Input{Industry index dataset $\boldsymbol{F}=\{F_1, F_2, \ldots, F_n\}$; the number of clusters $k$.}
    \Output{$k$ industry clusters.}
    Construct an adjacency matrix based on the similarity of the industry index\;
    
    Calculate the Laplacian matrix $\boldsymbol{L}=\boldsymbol{D}-\boldsymbol{A}$, where $\boldsymbol{D}=\operatorname{diag}(d_1, d_2, \ldots, d_n)$, $d_i=\sum_{j=1}^n A_{i j}, i=1,2, \ldots, n$\;
    
    Calculate the normalized Laplacian matrix $\tilde{L}=\boldsymbol{D}^{-1/2} \boldsymbol{L}\boldsymbol{D}^{1/2}$;
    Calculate the eigenvectors corresponding to the first $k$ smallest eigenvalues of the normalized Laplacian matrix $z_1, z_2, \ldots, z_k$\;
    
    Construct the matrix $\boldsymbol{Z}_{n \times k}$ combining $k$ _n_-dimensional feature vectors\;
    
    The matrix $\boldsymbol{Z}_{n \times k}$ is clustered into categories by the \textit{k}-means algorithms $C_1, C_2, \ldots, C_k$.
    
\end{algorithm}

In the SC algorithm, the number of clusters $k$ is pre-determined. In this study, we determine the number of clusters by the elbow methods. Based on the SC algorithm, the industry index matrix $II=\{F_1, F_2, \ldots, F_n\}$ is clustered into $k$ industry clusters $C_{1}, C_{2}, \ldots, C_{i}, \ldots, C_{k}$ where $C_{i}=\{F_{i}^{1}, F_{i}^{2}, \ldots, F_{i}^{n_{i}}\}$, where $n_i$ is the $n_i\text{-}th$ industry index of the $i\text{-}th$ cluster. For the $i\text{-}th$ cluster, the input vector is $\boldsymbol{x}_{i}=[x_{i}^{1}, x_{i}^{2}, \ldots, x_{i}^{n_{i}}]$ where $n_{i}\text{-}th$ industry index returns of the $i\text{-}th$ industry cluster.  

Based on the formed industry cluster, each industry cluster represents the highly correlated industry index within a period of time. Carefully selecting the most predictive industry cluster can provide crucial information that significantly enhances the prediction performance. The process of selecting important industry clusters can be viewed as a dynamic industry rotation effect, wherein we strategically identify and prioritize industry clusters that exhibit higher potential for providing valuable insights. The selection process empowers our model to capture and leverage the inherent patterns of the industry index data. Therefore, to extract the features of the industry rotation effect from the corresponding clusters, we utilize the multiple branch convolutional kernels. The details of the feature extraction process are shown from the Eq. (\ref{eq:convolutional operators}) to the Eq. (\ref{eq:output of MBCNN}). Based on the above steps, the prediction probability results $\boldsymbol{p}_{I}$ are obtained by the SC-MBCNN model.

\subsubsection{Feature extraction of the financial news}
The third data source in the proposed model is the financial news text. In reality, each news provider has a different level of authoritativeness and prediction ability for the target SMI movement. To capture the influence differentiation among multiple news providers, we employ the RNN-ER model to extract the intricate features embedded within the financial news text, which enables us to unravel the varying impacts and prediction strengths of different news providers. The steps of the feature extraction process are described as follows.

Suppose there are $m$ news providers corresponding to the $m$ datasets where each dataset includes the market data and the sentiment index of each news provider. The market data include the open price, the close price, the highest price, the lowest price, the trading volume, and the returns. The sentiment index of a news provider includes the positive sentiment index, the negative sentiment index, and the neutral sentiment index. Based on the prepared data, we first employ the sentiment dictionary-based method to calculate the sentiment polarity of a piece of financial news text \citep{pangOpinionMiningSentiment2008}. The dictionary-based sentiment classification method consists of the following three steps: 1) we carry out the text pre-processing steps which mainly include word tokenization, removal of punctuation and stopwords; 2) we utilize the word vector to represent the financial news where each value in the word vector is the weighted value of a word which is calculated by the word frequency-inverse document frequency (TF-IDF) method; 3) we transform the word vector into a sentiment vector based on the financial sentiment dictionary. And then, we adopt the sentiment index method in previous study \citep{gaoIntegratingSentimentsMultiple2022} to obtain the sentiment index of different news providers. As a result, combined with the historical market data, the input of the $i\text{-}th$ news provider is $\boldsymbol{x}_{i}=\{x_{i, 1}, x_{i, 2}, \ldots, x_{i, s}\}$.  

Based on the input $\boldsymbol{x}_{i}=\{x_{i, 1}, x_{i, 2}, \ldots, x_{i, s}\}$, we build a double layer recurrent neural network models as base classifier. The reason why we employ a double-layer RNN model is that the sentiment index of a news provider and the historical market data is the sequential data which is well handled by the RNN model. We briefly describe the feature extraction process from multiple news providers as follows. 

Given the input at the time step $t_1$ $\boldsymbol{x}_{i}^{t_{1}}=\{x_{i, 1}^{t_{1}}, x_{i, 2}^{t_{1}}, \ldots, x_{i, s}^{t_{1}}\}$, the output of the $j\text{-}th$ neuron in the first recurrent layer and the second recurrent layer is shown in the following equations: 
\begin{equation}
    h_{i, j}^{t_{1}(1)}=\theta(\boldsymbol{U} \cdot \boldsymbol{x}_{i}^{t_{1}}+\boldsymbol{W}^{(1)} \cdot \boldsymbol{h}_{i}^{t_{1}-1,(1)}+b_{i, j}^{(1)}),
\end{equation}
\begin{equation}
    h_{i, j}^{t_{1,1}(2)}=\theta(\boldsymbol{V} \cdot \boldsymbol{h}_{i}^{t_{1}(1)}+\boldsymbol{W}^{(2)} \cdot \boldsymbol{h}_{i}^{t_{1}-1,(2)}+b_{i, j}^{(2)}),
\end{equation}
where $\theta$ is the activation function, $b_{i, j}^{(1)}$ and $b_{i, j}^{(2)}$ are respectively the bias, $\boldsymbol{U}, \boldsymbol{V}, \boldsymbol{W}^{(1)}, \boldsymbol{W}^{(2)}$ are respectively the connection weights between the input layer and the first recurrent layer, the first recurrent layer and the second recurrent layer, the first recurrent layer at the time step $t_1-1$ and that at the time step $t_1$, the second recurrent layer at the time step $t_1-1$ and that at the time step $t_1$. And then, we employ the fully connected layer to obtain the output values $\boldsymbol{r}_{i}$ based on the $i\text{-}th$ news provider.

Based on $m$ prediction results of multiple news providers, we employ the evidential reasoning rule to integrate the above results. In the ER-based ensemble rule, each news provider is viewed as an evidence. The purpose of utilizing multiple evidence is to improve the prediction performance. The reason why we employ the ER rule is that it not only considers the weight of each evidence but also considers the reliability of each evidence. In the target SMI prediction task, it is crucial to take into account the reliability of evidence because not all evidence sources are equally reliable. The superiority and the calculation process of multiple news providers has been investigated in previous study \citep{gaoIntegratingSentimentsMultiple2022}. Based on RNN-ER model, we obtain the prediction probability results $\boldsymbol{p}_{M}$ provided by the financial news.

\subsection{Stage 2: Dynamic stacking ensemble model}
Utilizing the global SMIs data, industry index data, and financial news text data, the MBCNN model, SC-MBCNN model, and RNN-ER model are employed to generate the prediction probabilities, denoted as $\boldsymbol{p}_G$, $\boldsymbol{p}_I$, $\boldsymbol{p}_M$. These three sets of predicted probabilities collectively form the extracted features $\boldsymbol{x}=\{x_{1}, x_{2}, \ldots, x_{6}\}$ in which $x_1$ and $x_2$ represent the up and down prediction probabilities of the MBCNN model, $x_3$ and $x_4$ represent the up and down prediction probabilities of the SC-MBCNN model, $x_5$ and $x_6$ represents the up and down prediction probabilities of the RNN-ER model.

To enhance the prediction capabilities, it is imperative to integrate the extracted features obtained from heterogeneous classifiers. Common ensemble techniques include methods such as majority voting, plurality voting, weighted voting, and the stacking method. Among them, the stacking method has been identified as particularly suitable for the integration of heterogeneous classifiers \citep{wolpertStackedGeneralization1992,papouskovaTwostageConsumerCredit2019,cuiTwostageStackingHeterogeneous2021}. In a stacking-based ensemble framework, the base classifiers are generated in the first stage. And then, the meta-classifier is used to learn the results produced by these base classifiers.

For time series data, the interaction between extracted features and the target SMI movement exhibits dynamic characteristics, i.e., there exists distinct non-linear patterns between the extracted features and the target SMI movement. Hence, to effectively capture these temporal dynamics, the utilization of multiple meta-classifiers for feature integration is an effective strategy. Therefore, we design a dynamic stacking ensemble model for the integration of these extracted features. Specifically, we first utilize multiple meta-classifiers within the ensemble model. And then, we dynamically select the optimal meta-classifier for each time window. These meta-classifiers include LR, KNN, SVM-type model (the RBF function and polynomial function), ExtraTrees (ET) model, Random Forests (RF) model and the ANN model. In Appendix B, we provide a comprehensive description of the calculation process employed by the multiple meta-classifiers. Herein, we emphasize the selection process of the optimal meta-classifier as follows.

By employing the sliding time window splitting method in \citep{gevaEmpiricalEvaluationAutomated2014}, we partition the dataset into ten time windows, denoted as $D_{1}, D_{2}, \ldots, D_{10}$. Within each time window, the dataset is initially divided into a train set and a validation set. Subsequently, the train set is further divided into two segments: the first segment is utilized to extract features from the diverse data sources, while the second segment is employed to train the meta-classifier. It is important to note that the prediction capabilities of each meta-classifier may vary across time windows. To consider the characteristic, we dynamically select the meta-classifier with the most optimal prediction performance in each time window, where the accuracy metric is used as the evaluation metric.

Based on the generated set of meta-classifiers $M\text{-}C\text{-}L=\{M\text{-}C_{1}, M\text{-}C_{2}, \ldots, M\text{-}C_{7}\}$, we use the following equation to select the optimal meta-classifier of each time window: 
\begin{equation}
    M\text{-}C_{i}=\arg \max (\operatorname{Accuracy}(M\text{-}C_{i, j})), j=1,2, \ldots, 7.
\end{equation}

The chosen meta-classifier then integrates the generated features from the global SMI data, the industry index data, and the financial news data for forecasting the target SMI movement.

\subsection{Stage-by-stage optimization}
The proposed TDSE model includes two stages: Stage 1 focuses on extracting features from different data sources by integrating investor knowledge and suitable network structure; Stage 2 is the dynamic stacking ensemble model for the feature fusion and target SMI movement prediction, which combines the LR model, KNN model, RBF-SVM model, Poly-SVM model, RF model, ET model and ANN model.

As there are numerous of important hyper-parameters for both stages, it is crucial to optimize these hyper-parameters for a better generalization ability. To achieve this, we employ the optimization algorithm to fine-tune the important hyper-parameters in Stage 1 and Stage 2. Based on the optimization process, we can effectively identify the optimal hyper-parameter settings that maximize the prediction performance of the proposed model.

In Stage 1 of our proposed model, there are a total of 42 hyper-parameters that require optimization. Specifically, the MBCNN model consists of 8 hyper-parameters, the RNN-ER model has 26 hyper-parameters, and the SC-MBCNN model has 8 hyper-parameters. In Stage 2, there are a total of 11 hyper-parameters that need to be optimized, which includes the penalty value for the LR model, the penalty value and Gamma value for the RBF-SVM model, the penalty value and polynomial value for the Poly-SVM model, the number of decision trees for the RF model, the number of neurons in the first, second, and third layer for the ANN model, the number of neighborhood samples for the KNN model, and the number of decision trees for the ET model.

Optimizing all 53 hyper-parameters simultaneously can lead to a large solution space, resulting in reduced search efficiency and a higher risk of getting trapped in local optima. To address the problem, we employ a stage-by-stage optimization method to determine the hyper-parameters. Specifically, we first optimize the important hyper-parameters in Stage 1, which includes fine-tuning the hyper-parameters to achieve optimal performance for each feature extraction process of Stage 1. Once the hyper-parameters for Stage 1 are determined, we proceed to optimize the important hyper-parameters in Stage 2. The proposed stage-by-stage optimization approach can improve the search efficiency and reduce the risk of local optima.

During the optimization process across Stage 1 and Stage 2, a total of 53 hyper-parameters need to be optimized. Becasue this task involves optimizing numerous parameters, it is crucial to select an appropriate optimization algorithm. Common optimization algorithms include the Genetic Algorithm (GA), Particle Swarm Optimization (PSO), Grey Wolf Optimizer (GWO), Firefly Algorithm (FA), and Harmony Search (HS) Algorithm. Among these, the Genetic Algorithm (GA) is particularly noteworthy due to its robustness and effective search capabilities across large and complex spaces. The advantages of GA are numerous: (1) it maintains a diversity of solutions through its population-based approach; (2) it enables the exploration of multiple areas of the solution space simultaneously; (3) it prevents premature convergence to local optima.  Besides, its strong global search capability reduces the risk of stagnating at suboptimal solutions. Therefore, we employ the GA algorithm to optimize the aforementioned hyper-parameters. The details of the proposed TDSE-GA model are shown the Algorithm \ref{algo:TDSE-GA}.

\begin{algorithm}[H]
    \scriptsize{
    
    \SetKwInOut{Input}{Input}
    \SetKwInOut{Output}{Output}
    \caption{The TDSE-GA model}
    \label{algo:TDSE-GA}
    
    \Input{The population size $n$; the maximum generation number $H$; the stalled maximum generation number; the global SMI data $D_G$; the industry index data $D_I$; the financial news data $D_M$; the meta classifier $M\text{-}C$; the fitness function $fit()$. }
    \Output{The optimal classifier in Stage 1: $MBCNN_{L,b}$, $SC\text{-}MBCNN_{L,b}$, $RNN\text{-}ER_{L,b}$, the optimal classifier in Stage 2: $M\text{-}C\text{-}L_b$.}
    \BlankLine
    
    Divide the dataset $\boldsymbol{D}$ into ten sub-datasets using the time sliding window approach\;
    \For{$j=\{1,2, \ldots, 10\}$}{
    Parallel optimize feature extraction models in Stage 1: $\boldsymbol{p}_{G, j}=MBCNN(D_{G, j})$, $\boldsymbol{p}_{I,j}=SC\text{-}MBCNN(D_{I,j})$, $\boldsymbol{p}_{M,j}=RNN\text{-}ER(D_{M,j})$\;
    }
    $h \leftarrow 0$, $sh \leftarrow 0$\;
    Initialize the population in Stage 2: $P^{(h)} \leftarrow\{M\text{-}C\text{-}L_{1}^{(h)}, M\text{-}C\text{-}L_{2}^{(h)}, \ldots, M\text{-}C\text{-}L_{1}^{(h)}, \ldots, M\text{-}C\text{-}L_{n}^{(h)}\}$\;
    \For{$i=\{1,2, \ldots, n\}$}{
        \For{$j=\{1,2, \ldots, 10\}$}{
            \For{$m=\{1,2, \ldots, 7\}$}{
                Calculate the accuracy of the meta-classifier: $Accuracy_{j, m}(M\text{-}C_{j, m}^{(h)}(\boldsymbol{p}_{G, j}, \boldsymbol{p}_{I, j}, \boldsymbol{p}_{M, j}))$\;
            }
            Select the optimal meta-classifier:  $M\text{-}C\text{-}L_{i, j}^{(h)} \leftarrow \arg \max (Accuracy(M\text{-}C_{i, j, m}^{(h)})), m=1,2, \ldots, 7$\;
        }
    }
    Evaluate each individual for the initialized population: $fit_{i}^{(h)} \leftarrow fit(M\text{-}C\text{-}L_{i}^{(h)})$\;
    Select the individual with the best fitness value in the initialized population: $M\text{-}C\text{-}L_{b}^{(h)}$\;
    $M\text{-}C\text{-}L_{b} \leftarrow M\text{-}C\text{-}L_{b}^{(h)}$, $fit_{b} \leftarrow fit_{b}^{(h)}$\;
    \While{$h<H$}{
        Select elite individuals from the population $P^{(h)}$\;
        Generate the offspring individuals based on the crossover operation and the mutation operation\;
        Generate the new population $P^{(h+1)}$ based on the elite individuals and the offspring individuals\;
        Evaluate each individual in $P^{(h+1)}$: $fit_{i}^{(h+1)} \leftarrow fit(M\text{-}C\text{-}L_{i}^{(h+1)})$\;
        Select the individual with the best fitness value in $P^{(h+1)}$: $M\text{-}C\text{-}L_{b}^{(h+1)}$\;
        \eIf{$fit_{b}^{(h+1)}>fit_{b}$}{
            $M\text{-}C\text{-}L_{b} \leftarrow M\text{-}C\text{-}L_{b}^{(h+1)}$, $M\text{-}C\text{-}L_{b} \leftarrow M\text{-}C\text{-}L_{b}^{(h+1)}$, $sh \leftarrow 0$\;
            }
            {
            $sh \leftarrow sh+1$\;
            \lIf{$sh>SH$}{break}
            }
        $h \leftarrow h+1$;
    }
    }
\end{algorithm}

Given the global SMI data, the industry index data, the financial news data, Algorithm 2 finds the optimal solution. 

\textbf{Step 1: Parallel optimize the feature extraction model in Stage 1}

To begin, the dataset is divided into ten sub-datasets (Line 1). Subsequently, the MBCNN, SC-MBCNN, and RNN-ER model are optimized in parallel with each sub-dataset (Line 2-4). As a result, the extracted features in the $j\text{-}th$ sub-dataset are respective $\boldsymbol{p}_{G, j}$, $\boldsymbol{p}_{I, j}$, and $\boldsymbol{p}_{M, j}$.

\textbf{Step 2: Optimize the hyper-parameter in Stage 2}

In Line 5, the generation number and the stalled generation number in Stage 2 are initialized to 0. And the initialized population is $P^{(H)}$. Following that, the accuracy of meta-classifiers of each individual is calculated in each time period, which helps assess the performance of each meta-classifier (Line 10). To adapt to the data patterns across different time periods, we select the optimal meta-classifier for each specific time periods based on its accuracy (Line 12). And then, the fitness value of each individual is calculated by the following equation:
\begin{equation}
    fit_{i}^{(h)}=\frac{1}{K} \sum_{k=1}^{K} Accuracy_{i, k},
    \label{eq:fitness function}
\end{equation}
where $K=9$ means that we use the first nine sub-datasets to determine whether an individual is a superior selection. The individual with the best fitness value is selected as the current best individual (Line 16), while the current best individual is temporarily selected as the best individual (Line 17).

Line 18-32 describe the operations of the GA algorithm, i.e., the selection, crossover, and the mutation operations. First, the elite individuals with better fitness value are chosen from the current population. Based on the elite individuals, offspring individuals are generated by the crossover and mutation operations. And then, these individuals and offspring individuals are then combined to form a new population (Line 21). And then, the fitness value of each individual in the new population is evaluated by the Eq. (\ref{eq:fitness function}).

The process (Line 24-31) determines whether to replace the temporary best individual with the best individual in $P^{(h+1)}$ ($M\text{-}C\text{-}L_{b}^{(h+1)}$) based on their fitness values. If the fitness value of is greater than that of the temporary best individual, the temporary best individual is replaced. And the number of stalled generation is reset to 0. Otherwise, the temporary best individual remains unchanged, and the number of stalled generation increases by 1.

The iteration process (Line 18-31) is repeated until one of the following termination conditions is satisfied: (1) The TDSE-GA model reaches the maximum generation number; (2) The TDSE-GA model reaches the maximum stalled generation number. When the proposed model terminates, it output the optimal individual in Stage 1 and Stage 2, i.e., the optimal classifier in Stage 1: $MBCNN_{L,b}$, $SC\text{-}MBCNN_{L,b}$, $RNN\text{-}ER_{L,b}$, the optimal classifier in Stage 2: $M\text{-}C\text{-}L_{b}$. 

\section{Experimental study}
To verify the effectiveness of the proposed model, we apply it to predict the daily movement of Shanghai Securities Composite index, SZSE Component index and Growth Enterprise index in Chinese stock market, and compare its performance with other methods. Based on the proposed model, we design a trading strategy and compare it with other trading strategies in the terms of accumulated return and Sharpe ratio. The experimental results are shown as follows.

\subsection{Data description}
The target SMIs include the Shanghai Securities Composite index (SSEC), the Composite index of Shenzhen Stock Market (SZEC), and the Growth Enterprise Index (GEI). The data for these target SMIs comes from China Stock Market \& Accounting Research Database (CSMAR). We obtain global SMIs data from the investing.com website, which includes 30 major SMIs from Asia, Europe, and Americas. The industry index data is obtained from the Wind database, including 97 industry index data. We obtain the financial news text data from the Tushare platform, which is one of the largest financial big data open communities in China. We collect a total of 751,186 news texts from five financial news sources: Sina Finance, Wall Street CN, Straight flush, East money, and Yun CaiJing. The collected data spans from October 11, 2018 to April 28, 2021. The overview of the collected financial news texts is shown in Table \ref{table:brief information of news}. 

\begin{table}[!ht]
    \centering
    \setlength{\belowcaptionskip}{10pt}
    \caption{The brief information of the collected financial news text}
    \label{table:brief information of news}
    \small
    \begin{tabular}{cccc}
    \hline
        No. & News provider & Number of news text & Percentage (\%)  \\ 
    \hline
        1 & Sina Finance & 255,225 & 33.98   \\ 
        2 & Wall Street CN & 67,962 & 9.05  \\ 
        3 & Straight flush & 117,206 & 15.60  \\ 
        4 & East money & 140,462 & 18.70  \\ 
        5 & Yun CaiJing & 170,331 & 22.67  \\ 
          & Total & 751,186 & 100.00  \\ 
    \hline
    \end{tabular}
\end{table}

For the text data collected from various financial news sources, we remove punctuations, numbers and stop words of each piece of financial news, and tokenize the remaining words. After the preprocessing step, we utilize the sentiment index calculation method in \citep{gaoIntegratingSentimentsMultiple2022} to determine the sentiment indexes of each news provider. In the calculation process, we employ the Chinese sentiment dictionary \citep{JiangMeiTiWenBenQingXuYuGuPiaoHuiBaoYuCe2021}, which consists of 5,890 positive words and 3,338 negative words. The statistical results on sentiment indexes and news volume across different financial news providers are presented in Table C.3 in the Appendix.

\subsection{Evaluation metrics}
We utilize the time sliding window approach introduced in \citep{gevaEmpiricalEvaluationAutomated2014} to divide the dataset into a train set and a test set. Specifically, we set the number of sliding windows to 10. Given that the dataset spans 31 months, in each time window, the initial 11 months' data is used as the train set, while the subsequent three months' data is utilized as the test set.
We aim to forecast the target SMI movement, which is a classification task. Therefore, we use the common evaluation metrics used in classification task, including accuracy, precision, recall, \textit{F}-measure, and AUC value. The classification results are summarized in the confusion matrix presented in Table \ref{table:confusion matrix}.

\begin{table}[!ht]
\centering
\setlength{\belowcaptionskip}{10pt}
\caption{Confusion matrix of the classification task}
\small
\label{table:confusion matrix}
\begin{tabular}{lll}
    \hline
    \multirow{2}{*}{Actual result} & \multicolumn{2}{c}{Predicted result} \\ 
      & Up & Down  \\ 
    \hline
    Up & TU & FD  \\ 
    Down & FU & TD  \\ 
    \hline
\end{tabular}
\end{table}

In Table \ref{table:confusion matrix}, TU (True Up) indicates that both the predicted result and the actual result are 'Up'. TD (True Down) indicates that both the predicted result and the actual result are 'Down'. If the predicted result is 'Down' while the actual result is 'Up', it is labeled as FD (False Down). If the predicted result is 'Up' while the actual result is 'Down', it is labeled as FU (False Up). The accuracy, precision, recall and \textit{F}-measure metrics are calculated using the following equations:
\begin{equation}
    Accuracy =\frac{TU+TD}{TU+TD+FU+FD},
\end{equation}

\begin{equation}
    Precision =\frac{TU}{TU+FU},
\end{equation}

\begin{equation}
    Recall =\frac{TU}{TU+FD},
\end{equation}

\begin{equation}
    F\text{-}measure =\frac{2 \times Precision \times Recall}{Precision+Recall}.
\end{equation}

Area Under Curve (AUC) is the area under the ROC curve where the FPR is the horizontal axis and the TPR is the vertical axis. The FPR and TPR are calculated using the following equations:
\begin{equation}
    FPR=\frac{FU}{FU+TD},
\end{equation}

\begin{equation}
    TPR=\frac{TU}{TU+FD}.
\end{equation}

The AUC value measures the overall performance of a classifier across all possible classification thresholds, primarily used to assess the classifier's generalization ability. A higher AUC value indicates better classification performance, with a value closer to 1 indicating greater performance in distinguishing between positive and negative samples.

\subsection{Experimental setting}
To validate the effectiveness of the proposed model, we have designed four experiments in this study. The first experiment aims to compare the prediction ability of data sources for forecasting the target SMI movement. In this experiment, we employ the following models as baselines: Artificial Neural Network (ANN), Support Vector Machine (SVM), Random Forest (RF), Convolutional Neural Network (CNN), Recurrent Neural Network (RNN), Long Short-Term Memory (LSTM), Gate Recurrent Unit (GRU), and 2-Dimension Convolutional Neural Network (2D-CNN). The inputs for the above models include market data, global SMI data, industry index data, and financial news text.

The second experiment focuses on verifying the effectiveness of the proposed TDSE-GA model, which takes into account the financial data patterns of diverse data sources. These comparison models include the Random Prediction model, the Label \textit{t}-1 model, the MBCNN-GA model \citep{gaoForecastingOvernightReturn2022}, the SC-MBCNN-GA model, the RNN-ER-GA model \citep{gaoIntegratingSentimentsMultiple2022}, the ANN+PCA model \citep{maMultisourceAggregatedClassification2022}, the MKL model \citep{shynkevichForecastingMovementsHealthcare2016}, the FCN model \citep{wangTimeSeriesClassification2017}, and the Transformer model \citep{vaswaniAttentionAllYou2017}. For the MBCNN-GA, SC-MBCNN-GA, and RNN-ER-GA models, the inputs are global SMIs data, industry index data and financial news text, respectively. For the MKL, ANN+PCA, FCN, Transformer, and the proposed TDSE-GA model, the inputs include global SMIs data, industry index data, and financial news text. We set the lag period for global SMIs and financial news text at 1, and the lag period for industry index data at 5. This setting is consistent with the previous studies \citep{,gaoIntegratingSentimentsMultiple2022,gaoForecastingOvernightReturn2022}.

In the third experiment, we assess the economic results based on different trading strategies. To investigate whether the proposed TDSE-GA model can provide excess returns, we design a trading strategy based on this model and compare it with two other types of trading strategies: simple trading strategies and trading strategies based on models used in related studies, i.e., MKL, ANN+PCA, FCN, Transformer models. 

The fourth experiment focuses on the dynamic selection process of meta-classifiers. Since the proposed model enhances prediction performance by progressively selecting the optimal classifier across different time periods, this experiment is designed to examine the robustness of the model and the dynamic selection process of the meta-classifiers.

\subsection{Experimental results}
\subsubsection{Prediction results of different data sources}
\begin{figure}
    \centering
    \includegraphics[width=1\linewidth]{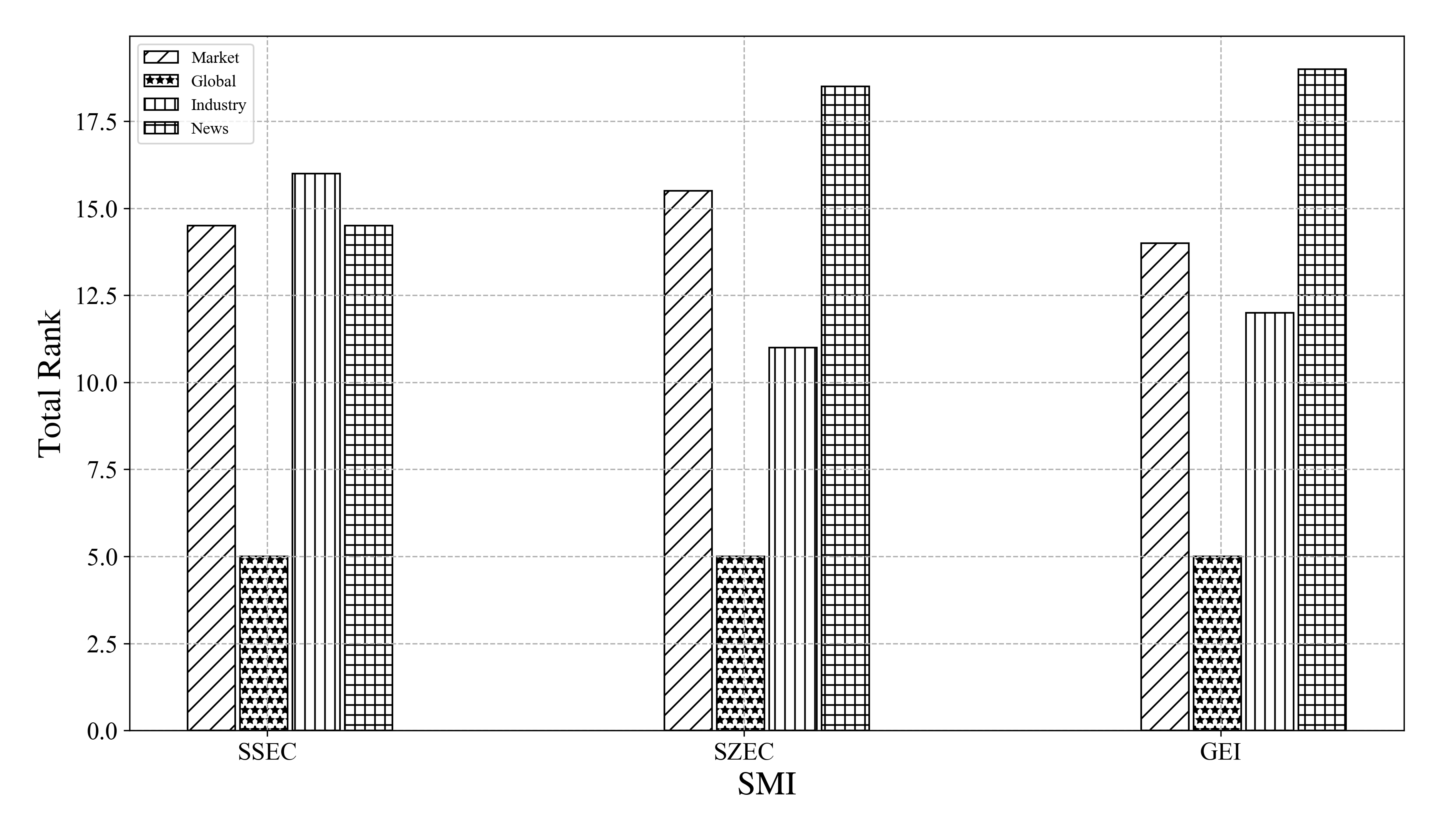}
    \caption{Total Rank results for different data sources on SSEC, SZEC, and GEI datasets}
    \label{fig:rank of different data sources}
\end{figure}

In this experiment, we compare the prediction results of different data sources, namely the market data, the global SMIs data, industry index data, and financial news text. The purpose of this experiment is to investigate whether adding additional data sources alongside the historical market data can improve the prediction capacities of forecasting models. To provide a comprehensive overview of the predictive capabilities of different data sources on the SSEC, SZEC, and GEI datasets, we introduce the Total Rank metric. This metric is defined as the average "Rank" of each data source. The "Rank" metric represents the ranking of different data sources based on their average predictive performance across various machine learning methods for each evaluation metric (accuracy, recall, precision, \textit{F}-measure, and AUC). The Total Rank value serves as a key indicator of predictive performance, with a lower value indicating superior performance of a data source. The Total Rank results for different data sources on the SSEC, SZEC, and GEI datasets are shown in Fig. \ref{fig:rank of different data sources}. The detailed calculation process and results are presented in Section C.2 of the Appendix.

From the results shown in Fig. \ref{fig:rank of different data sources}, we first observe that integrating global SMIs data with the historical market data remarkably outperforms other data sources, showcasing the best Total Rank value (5) on the SSEC, SZEC, and GEI datasets. Furthermore, we find a discernible enhancement in prediction performance upon the incorporation of industry index data within the SZEC and GEI datasets. Conversely, the inclusion of news data demonstrates an adverse impact on prediction performance within both the SZEC and GEI datasets. Last, we find that the supplementation of either industry index data or news data yields no obvious improvement in prediction performance in the SSEC dataset.

\subsubsection{The prediction results comparison of different prediction models}
The experimental results presented in Section 4.4.1 reveal distinct prediction capabilities among the global SMIs data, industry index data, and financial news text. Specifically, integrating historical market data with global SMIs data leads to an obvious improvement in prediction performance. In contrast, the inclusion of industry index data or financial news data does not provide substantial improvements in forecasting capabilities. Therefore, to fully harness the prediction potential of the above data sources and enhance the prediction performance, we propose the TDSE-GA model specifically designed to integrate multi-modal data and account for the financial data patterns from diverse data sources. To verify the effectiveness of the proposed model, we compare it with other competing methods. The details of the proposed model and the competing methods are shown in Table \ref{table:description of model}. To be consistent with the parameter settings in related studies \citep{gaoIntegratingSentimentsMultiple2022}, we determine the hyper-parameters for the proposed model, as presented in the Table C.3 of the Appendix.

\begin{table}[!ht]
    \centering
    \caption{Description of the proposed model and competing methods}
    \label{table:description of model}
    \footnotesize
    \begin{tabularx}{\textwidth}{m{3cm}lX}
        \hline
        Type & Prediction Model & Description  \\ 
        \hline
        \multirow{2}{*}[-0.2cm]{Simple model}
        & Random Prediction & The predicted value is determined by random selection  \\ 
        ~ & Label \textit{t}-1 & The predicted value is determined by the label at period \textit{t}-1  \\ \hline
        
        \multirow{3}{*}[-0cm]{\parbox{3cm}{Prediction model based on single data source}} 
        & MBCNN-GA & The model input is the global SMIs data  \\ 
        ~ & SC-MBCNN-GA  & The model input is the industry index data  \\ 
        ~ & RNN-ER-GA  & The model input is the financial news text  \\ \hline
        
        \multirow{4}{*}[-0cm]{\parbox{3cm}{Multi-source data fusion model}} 
        & MKL  & \multirow{4}{*}{\parbox{6.5cm}{The model input includes the global SMIs data, the industry index data and the financial news text}}  \\
        ~ & ANN+PCA &   \\ 
        ~ & FCN  &   \\ 
        ~ & Transformer  &   \\ \hline
        The proposed model & TDSE-GA & The model input includes the global SMIs data, the industry index data and the financial news text  \\ \hline
    \end{tabularx}
\end{table}

Table \ref{table:prediction results of different models} shows the prediction results of the proposed model and the competing methods on the SSEC, SZEC, and GEI datasets. The results are averaged over 10 time windows to provide a comprehensive evaluation. In Table \ref{table:prediction results of different models}, the numbers in boldface within each column represent the highest values for a specific SMI and evaluation metric. The improvement rate refers to the ratio by which the proposed TDSE-GA model outperforms the best results of the comparison models. Based on the prediction results in Table 8, we can draw the following key observations. (1) The proposed TDSE-GA model is superior to other competing methods in terms of the accuracy (SSEC, SZEC, and GEI), recall (SSEC), precision (SSEC, SZEC, and GEI), \textit{F}-measure (SSEC, SZEC, and GEI) and AUC (SSEC, SZEC, and GEI). In terms of the Accuracy metric which is the most important indicator in SMI prediction tasks, the proposed TDSE-GA model has achieved improvements of 11.42\%, 7.94\%, and 7.73\% over the best comparison models for the SSEC, SZEC, and GEI indices, respectively. (2) The Label t-1 model is better than the Random Prediction method with respect to the prediction results on SSEC, SZEC, and GEI datasets. (3) Among models utilizing different data sources, the MBCNN-GA model demonstrates superior accuracy compared to the SC-MBCNN-GA model and the RNN-ER-GA model. (4) Regrading models integrating multi-modal data, the Transformer model delivers better results on SSEC and GEI datasets, while the ANN+PCA model outperforms other models on the SZEC dataset. (5) Models considering the financial pattern of individual data source generally provide better prediction results than those simply integrating multi-source data in a straightforward manner. For instance, the MBCNN-GA model achieves superior prediction results than the Transformer model on SSEC, SZEC, and GEI datasets.

\begin{table}[!ht]
    \centering
    \caption{The prediction results of different methods}
    \label{table:prediction results of different models}
    
    \begin{tabular}
        {
            p{2cm}
            llllll
        }
        \hline
        Method type & Model & Accuracy & Recall & Precision & F-measure & AUC  \\ \hline
        
        & \multicolumn{6}{c}{SSEC}\\
        \cmidrule(lr){2-7}
        
        \multirow{2}{*}[-0.0cm]{Simple model} & Random Prediction & 0.4657 & 0.4890 & 0.4488 & 0.4619 & 0.4687  \\
        ~ & Label \textit{t}-1 & 0.5255 & 0.4905 & 0.4919 & 0.4911 & 0.5123  \\
        \cmidrule(lr){2-7}
        
        \multirow{3}{*}[-0.0cm]{\makecell[l]{Single data \\ source model}}
         & MBCNN-GA & 0.5790 & 0.6112 & 0.5537 & 0.5808 & 0.5800  \\
         & SC-MBCNN-GA & 0.5573 & 0.6087 & 0.5367 & 0.5732 & 0.5565  \\
         & RNN-ER-GA & 0.5561 & 0.5058 & 0.5649 & 0.4929 & 0.5468  \\
        \cmidrule(lr){2-7}
        
        \multirow{5}{*}[-0.0cm]{\makecell[l]{Multi-source data \\ fusion model}}
         & ANN+PCA & 0.5112 & 0.5438 & 0.5029 & 0.5171 & 0.5021  \\ 
         & MKL & 0.4885 & 0.4369 & 0.2852 & 0.3382 & 0.4220  \\ 
         & FCN & 0.5134 & 0.5844 & 0.5054 & 0.5343 & 0.5204  \\ 
         & Transformer & 0.5511 & 0.5659 & 0.5549 & 0.5439 & 0.5558  \\ 
         & TDSE-GA & \textbf{0.6451} & \textbf{0.7236} & \textbf{0.6442} & \textbf{0.6521} & \textbf{0.6453}  \\ 
         & Improvement rate(\%) & 11.42 & 18.39 & 14.04 & 12.28 & 11.26  \\ \hline

        & \multicolumn{6}{c}{SZEC}\\
        \cmidrule(lr){2-7}
        
        \multirow{2}{*}[-0.0cm]{Simple model} & Random Prediction & 0.4587 & 0.4915 & 0.4836 & 0.4781 & 0.4619  \\ 
        ~ & Label \textit{t}-1 & 0.5399 & 0.5484 & 0.5456 & 0.5468 & 0.5282  \\
        \cmidrule(lr){2-7}
        
        \multirow{3}{*}[-0.0cm]{\makecell[l]{Single data \\ source model}} & MBCNN-GA & 0.5767 & 0.6268 & 0.5868 & 0.6068 & 0.5776  \\ 
        & SC-MBCNN-GA & 0.5703 & 0.6104 & 0.5800 & 0.5865 & 0.5668  \\ 
        ~ & RNN-ER-GA & 0.5696 & \textbf{0.8172} & 0.5614 & 0.6194 & 0.5398  \\
        \cmidrule(lr){2-7}
        
        \multirow{5}{*}[-0.0cm]{\makecell[l]{Multi-source data \\ fusion model}} & ANN+PCA & 0.5415 & 0.5519 & 0.5682 & 0.5518 & 0.5536  \\ 
        & MKL & 0.5145 & 0.4411 & 0.3193 & 0.3621 & 0.4584  \\ 
        ~ & FCN & 0.5219 & 0.5786 & 0.5414 & 0.5506 & 0.5209  \\ 
        ~ & Transformer & 0.5223 & 0.5401 & 0.5551 & 0.5345 & 0.5383  \\ 
        ~ & TDSE-GA & \textbf{0.6225} & 0.7708 & \textbf{0.6312} & \textbf{0.6624} & \textbf{0.6192}  \\ 
        ~ & Improvement rate(\%) & 7.94 & -5.68 & 7.57 & 6.94 & 7.20  \\ \hline

        & \multicolumn{6}{c}{GEI}\\
        \cmidrule(lr){2-7}
        
        \multirow{2}{*}[-0.0cm]{Simple model} & Random Prediction & 0.4618 & 0.4986 & 0.5217 & 0.4992 & 0.4668  \\ 
        ~ & Label \textit{t}-1 & 0.5035 & 0.5373 & 0.5373 & 0.5372 & 0.4888  \\
        \cmidrule(lr){2-7}
        
        \multirow{3}{*}[-0.0cm]{\makecell[l]{Single data 
        \\ source model}} & MBCNN-GA & 0.5809 & 0.7035 & 0.6088 & 0.6418 & 0.5718  \\ 
        & SC-MBCNN-GA & 0.5706 & 0.6436 & 0.6170 & 0.6162 & 0.5745  \\ 
        ~ & RNN-ER-GA & 0.5807 & \textbf{0.7533} & 0.5781 & 0.6122 & 0.5374  \\
        \cmidrule(lr){2-7}
        
        \multirow{5}{*}[-0.0cm]{\makecell[l]{Multi-source data \\ fusion model}}
        & ANN+PCA & 0.5546 & 0.5746 & 0.6078 & 0.5847 & 0.5259  \\ 
        & MKL & 0.5048 & 0.6525 & 0.4381 & 0.5154 & 0.4124  \\ 
        ~ & FCN & 0.5512 & 0.5817 & 0.6029 & 0.5851 & 0.5343  \\ 
        ~ & Transformer & 0.5579 & 0.5987 & 0.6114 & 0.5938 & 0.5958  \\ 
        ~ & TDSE-GA & \textbf{0.6258} & 0.7419 & \textbf{0.6823} & \textbf{0.6663} & \textbf{0.6188}  \\ 
        ~ & Improvement rate (\%) & 7.73 & -1.51 & 10.58 & 3.82 & 3.86  \\ \hline
    \end{tabular}
\end{table}

To test the significance difference level between the prediction results of the proposed model and the competing methods, we carry out the paired \textit{t}-tests. The results of the paired \textit{t} tests between the proposed model and the competing methods on the SSEC, SZEC, and GEI datasets are shown in Table \ref{table:Paired t tests}. Based on the results presented in Table \ref{table:Paired t tests}, we observe significant differences between the TDSE-GA model and other competing methods in terms of the accuracy, recall, precision, \textit{F}-measure, and AUC metrics on the SSEC dataset (except for the precision metric when compared to the Transformer model). The significance levels for the accuracy and AUC metrics are low as 0.05, while the significance level for the \textit{F}-measure metric is 0.01. These findings indicate the proposed TDSE-GA model effectively improves the prediction performance of the target SMI movement by taking account into the financial data patterns of multi-modal data. 

\begin{table}[!ht]
    \centering
    \caption{Paired \textit{t} tests between the proposed model and the competing methods}
    \label{table:Paired t tests}
    \begin{tabular}{llllllp{1cm}}
    \hline
    Method type & Model & Accuracy & Recall & Precision & F-measure & AUC  \\ \hline

    & \multicolumn{6}{c}{SSEC}\\
    \cmidrule(lr){2-7}
    
    \multirow{2}{*}[-0.0cm]{Simple model} & Random Prediction & 0.001\tnote{***} & 0.003\tnote{***} & 0.003\tnote{***} & 0.001\tnote{***} & 0.001\tnote{***}  \\ 
    ~ & Label \textit{t}-1 & 0.010\tnote{***} & 0.010\tnote{***} & 0.002\tnote{***} & 0.006\tnote{***} & 0.004\tnote{***}  \\
    \cmidrule(lr){2-7}

    \multirow{3}{*}[-0.0cm]{\makecell[l]{Single data \\ source model}}
    & MBCNN-GA & 0.032\tnote{**} & 0.032\tnote{**} & 0.028\tnote{**} & 0.004\tnote{***} & 0.009\tnote{***}  \\ 
    ~ & SC-MBCNN-GA & 0.000\tnote{***} & 0.024\tnote{**} & 0.001\tnote{***} & 0.002\tnote{***} & 0.000\tnote{***}  \\ 
    ~ & RNN-ER-GA & 0.004\tnote{***} & 0.046\tnote{**} & 0.042\tnote{**} & 0.006\tnote{***} & 0.000\tnote{***}  \\
    \cmidrule(lr){2-7}

    \multirow{4}{*}[-0.0cm]{\makecell[l]{Multi-source data \\ fusion model}}
    & ANN+PCA & 0.001\tnote{***} & 0.007\tnote{***} & 0.004\tnote{***} & 0.000\tnote{***} & 0.012\tnote{**}  \\
    ~ & MKL & 0.004\tnote{***} & 0.033\tnote{**} & 0.001\tnote{***} & 0.004\tnote{***} & 0.001\tnote{***}  \\ 
    ~ & FCN & 0.001\tnote{***} & 0.058\tnote{*} & 0.004\tnote{***} & 0.004\tnote{***} & 0.009\tnote{***}  \\
    ~ & Transformer & 0.043\tnote{**} & 0.008\tnote{***} & 0.122 & 0.006\tnote{***} & 0.049\tnote{**}  \\ \hline

    & \multicolumn{6}{c}{SZEC}\\
    \cmidrule(lr){2-7}
    
    \multirow{2}{*}[-0.0cm]{Simple model} & Random Prediction & 0.000\tnote{***} & 0.002\tnote{***} & 0.000\tnote{***} & 0.001\tnote{***} & 0.000\tnote{***}  \\ 
    ~ & Label \textit{t}-1 & 0.005\tnote{***} & 0.000\tnote{***} & 0.010\tnote{***} & 0.001\tnote{***} & 0.003\tnote{***}  \\
    \cmidrule(lr){2-7}
    
    \multirow{3}{*}[-0.0cm]{\makecell[l]{Single data \\ source model}}
    & MBCNN-GA & 0.043\tnote{**} & 0.001\tnote{***} & 0.079* & 0.003\tnote{***} & 0.049\tnote{**}  \\ 
    ~ & SC-MBCNN-GA & 0.029\tnote{**} & 0.007\tnote{***} & 0.101 & 0.015\tnote{**} & 0.050\tnote{**}  \\
    ~ & RNN-ER-GA & 0.106 & 0.613 & 0.082\tnote{*} & 0.348 & 0.006\tnote{***}  \\
    \cmidrule(lr){2-7}
    
    \multirow{4}{*}[-0.0cm]{\makecell[l]{Multi-source data \\ fusion model}} 
    & ANN+PCA & 0.064\tnote{*} & 0.002\tnote{***} & 0.126 & 0.017\tnote{**} & 0.161  \\
    ~ & MKL & 0.002\tnote{***} & 0.023\tnote{**} & 0.002\tnote{***} & 0.006\tnote{***} & 0.001\tnote{***}  \\ 
    ~ & FCN & 0.006\tnote{***} & 0.010\tnote{***} & 0.027\tnote{**} & 0.009\tnote{***} & 0.003\tnote{***}  \\
    ~ & Transformer & 0.000\tnote{***} & 0.001\tnote{***} & 0.026\tnote{**} & 0.001\tnote{***} & 0.008\tnote{***}  \\ \hline

    & \multicolumn{6}{c}{GEI}\\
    \cmidrule(lr){2-7}
    
    \multirow{2}{*}[-0.0cm]{Simple model} & Random Prediction & 0.001\tnote{***} & 0.009\tnote{***} & 0.003\tnote{***} & 0.001\tnote{***} & 0.003\tnote{***}  \\ 
    ~ & Label \textit{t}-1 & 0.000\tnote{***} & 0.005\tnote{***} & 0.001\tnote{***} & 0.001\tnote{***} & 0.000\tnote{***}  \\
    \cmidrule(lr){2-7}
    
    \multirow{3}{*}[-0.0cm]{\makecell[l]{Single data \\ source model}}
    & MBCNN-GA & 0.099\tnote{*} & 0.553 & 0.028\tnote{**} & 0.420 & 0.109  \\ 
    ~ & SC-MBCNN-GA & 0.073\tnote{*} & 0.101 & 0.046\tnote{**} & 0.080\tnote{*} & 0.115  \\ 
    ~ & RNN-ER-GA & 0.209 & 0.882 & 0.006\tnote{***} & 0.366 & 0.013\tnote{**}  \\
    \cmidrule(lr){2-7}
    
    \multirow{4}{*}[-0.0cm]{\makecell[l]{Multi-source data \\ fusion model}} 
    & ANN+PCA & 0.047\tnote{**} & 0.023\tnote{**} & 0.017\tnote{**} & 0.055\tnote{*} & 0.031\tnote{**}  \\ 
    ~ & MKL & 0.002\tnote{***} & 0.555 & 0.010\tnote{***} & 0.141 & 0.000\tnote{***}  \\ 
    ~ & FCN & 0.111 & 0.037\tnote{**} & 0.107 & 0.057\tnote{*} & 0.071\tnote{*}  \\ 
    ~ & Transformer & 0.115 & 0.054\tnote{*} & 0.109 & 0.056\tnote{*} & 0.610  \\ \hline
    \end{tabular}
    
    \begin{tablenotes}
        \item[*] statistically significant at the 0.1 level.
        \item[**] statistically significant at the 0.05 level.
        \item[***] statistically significant at the 0.01 level.
    \end{tablenotes}

\end{table}

\subsubsection{Economic results based on different trading strategies}
To assess the economic value of our proposed TDSE-GA model, we design a trading strategy based on its prediction signals. We then compare this designed strategy with other approaches: the naïve trading strategy and the trading strategy based on the predicted signals of related studies. In this experiment, we do not take into account transaction costs and taxes. The details of these three types of trading strategies are described as follows.

(1) Long trading strategy based on the predicted signals of the TDSE-GA model (TDSE-GA): if the TDSE-GA model predicts an upward movement, the trading system buys or holds the target investment product; if the prediction signals a downward movement, the trading system sells all positions of the target investment product and holds cash.

(2) The naïve trading strategies include the Buy \& Hold trading strategy and the long trading strategy based on random prediction. The Buy \& Hold trading strategy involves buying the target investment product on the first trading day and holding it until the last trading day. The long trading strategy based on random prediction follows the same principles as the TDSE-GA trading strategy but uses random prediction to generate the signals.

(3) Long trading strategy based on predicted signals of the model in related studies. The trading strategy shares principles with the TDSE-GA trading strategy but differs in the models used to generate predicted signals. The predicted signals are generated by the models in related studies, including the MKL model \citep{shynkevichForecastingMovementsHealthcare2016}, the ANN+PCA model \citep{zhongForecastingDailyStock2017}, the FCN model \citep{wangTimeSeriesClassification2017}, and the Transformer model \citep{vaswaniAttentionAllYou2017}.

\begin{figure}
    \centering
    \includegraphics[width=1\linewidth]{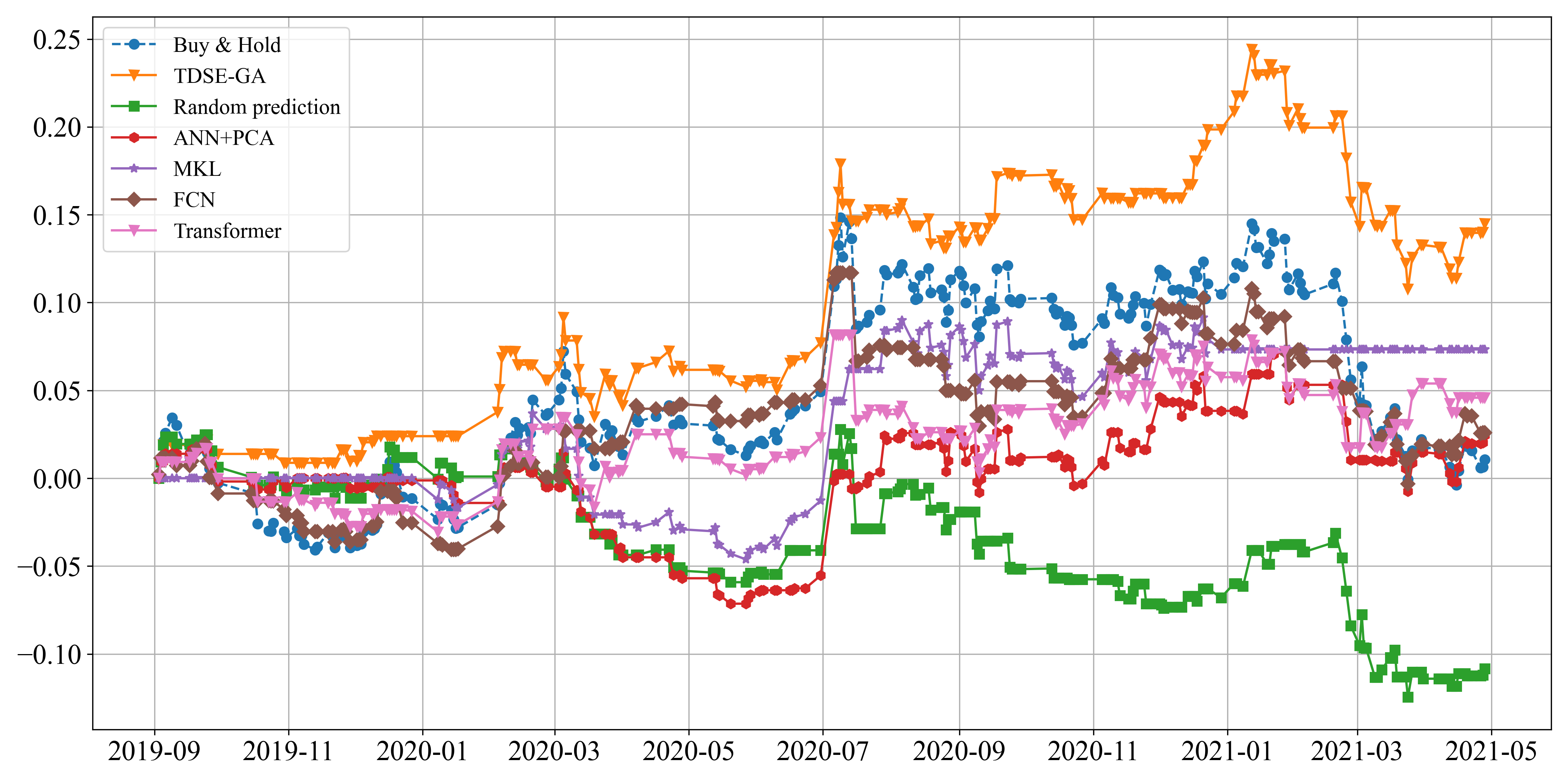}
    \caption{Accumulative return curve of different strategies on SSEC}
    \label{fig:SSEC return}
\end{figure}

\begin{figure}
    \centering
    \includegraphics[width=1\linewidth]{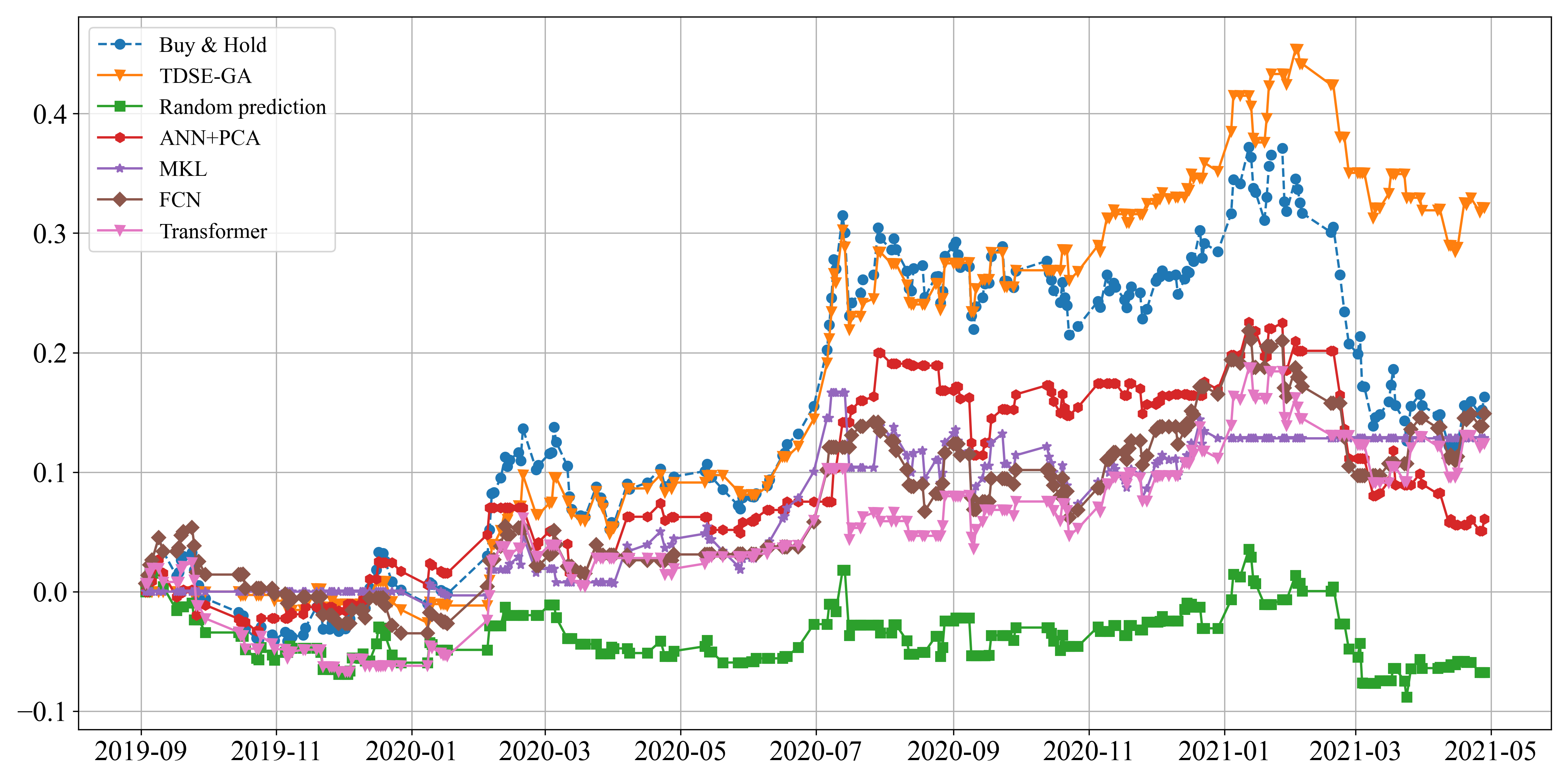}
    \caption{Accumulative return curve of different strategies on SZEC}
    \label{fig:SZEC return}
\end{figure}

\begin{figure}
    \centering
    \includegraphics[width=1\linewidth]{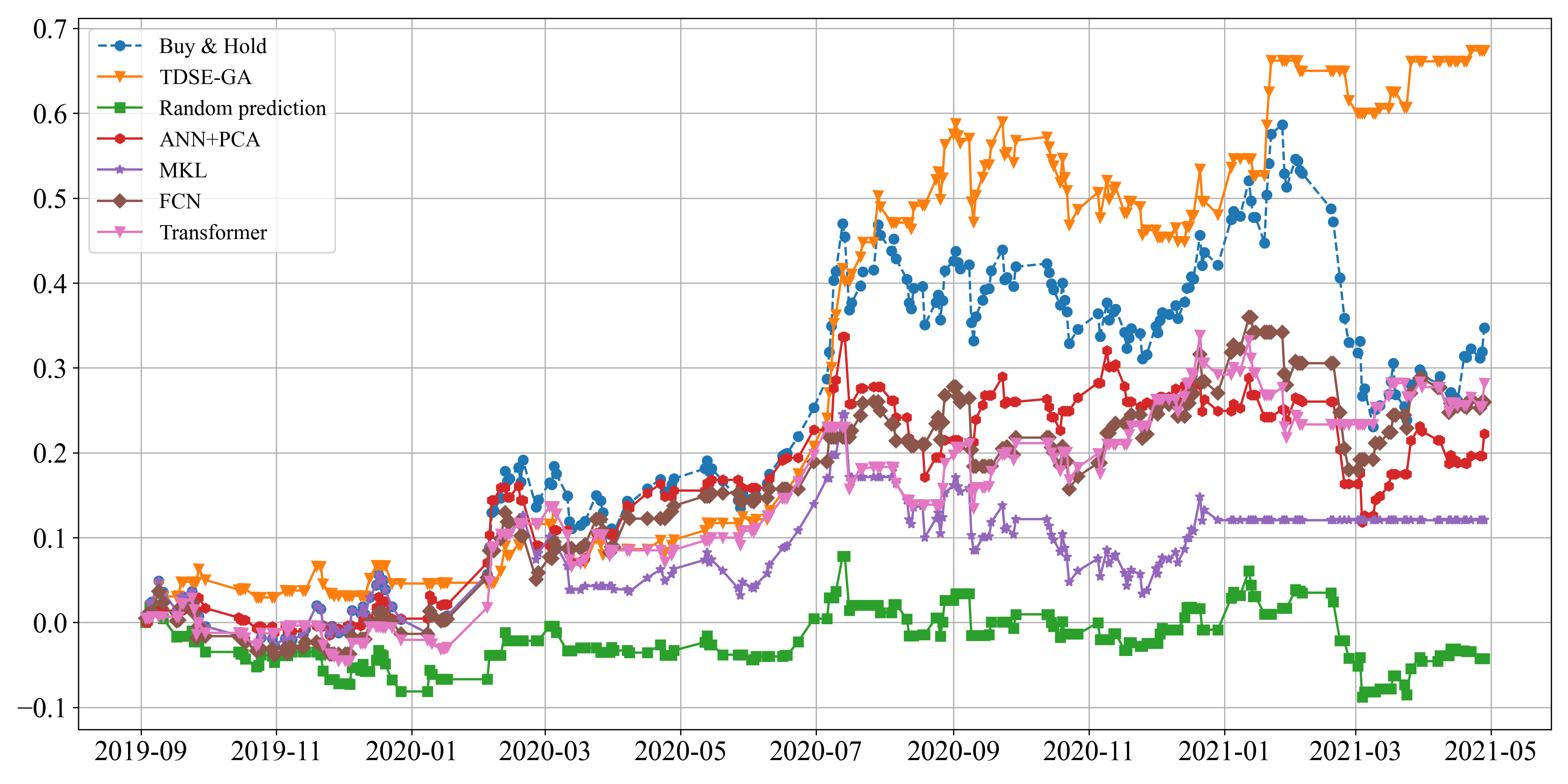}
    \caption{Accumulative return curve of different strategies on GEI}
    \label{fig:GEI return}
\end{figure}

The time period for these trading strategies is from September 1st, 2019 to April 30th, 2021. The accumulative return curves of different strategies on SSEC, SZEC, and GEI datasets are shown in Fig. \ref{fig:SSEC return}, Fig. \ref{fig:SZEC return}, and Fig. \ref{fig:GEI return}, respectively. Based on the analysis results, we make the following observations. First, the accumulative returns curves of the other trading strategies revolve around the curve of the Buy \& Hold strategy. This phenomenon is attributed to the return being derived from the correct predicted signal, which is directly related to the target SMI. Consequently, the Buy \& Hold trading strategy aptly reflects the earning results of the target SMI. Second, the trading strategy based on the proposed model consistently generates higher accumulative returns compared to the other trading strategies across the SSEC, SZEC, and GEI datasets. This outcome underscores the significant economic value of the proposed model in enhancing the prediction performance. Third, among the competing trading strategies, the strategy based on the random prediction yields the lowest accumulative return. In contrast, the trading strategy based on the Transformer model and the Buy \& Hold strategy outperform the others, yielding better accumulative returns on the SSEC, SZEC, GEI datasets.

To comprehensively evaluate the performance of various trading strategies, we calculate the accumulative returns, monthly returns, Sharpe ratio, daily returns, and maximum drawdown rate of each strategy. The comparative results of different trading strategies on the SSEC, SZEC, and GEI datasets are presented in Table \ref{table:Economic results} where the number in boldface within each column indicates the best result for that specific SMI. The improvement rate refers to the enhancement in performance achieved by a trading strategy based on the proposed model, as compared to the best competing strategy, under a specific evaluation metric. An interesting observation from Table \ref{table:Economic results} is that monthly returns and daily returns of trading strategies based on the ANN+PCA model, the FCN model, and the Transformer model show inconsistent directions. This disparity is due to the calculation methods: monthly returns consider the difference between the first and last trading days within a month, while daily returns represent the average of cumulative returns.

\begin{table}[!ht]
    \centering
    \caption{Results comparison of different trading strategies on the SSEC, SZEC, and GEI}
    \label{table:Economic results}
    \footnotesize 
    \begin{tabularx}{\textwidth}{
        l
        l
        >{\centering\arraybackslash}X
        >{\centering\arraybackslash}X
        >{\centering\arraybackslash}X
        >{\centering\arraybackslash}X
        >{\centering\arraybackslash}X
    }
    \hline
    SMI & Trading strategy & Accumulative returns & Monthly returns & Sharpe ratio & Daily returns & Maximum drawdown rate  \\ \hline
    \multirow{8}{*}{SSEC} 
    & Buy \& Hold & 0.01063 & 0.00390 & 0.17326 & 0.00004 & 0.13246  \\
    ~ & Random Prediction & -0.10826 & -0.00599 & -0.34054 & -0.00044 & 0.14822  \\ 
    ~ & ANN+PCA & 0.02447 & -0.00261 & -0.18592 & 0.00010 & 0.08644  \\ 
    ~ & MKL & 0.07331 & 0.00237 & 0.10780 & 0.00030 & 0.07994  \\
    ~ & FCN & 0.02601 & -0.00076 & -0.03984 & 0.00011 & 0.10765  \\ 
    ~ & Transformer & 0.04584 & -0.00299 & -0.14398 & 0.00019 & \textbf{0.07263}  \\ 
    & TDSE-GA & \textbf{0.14492} & \textbf{0.00489} & \textbf{0.40594} & \textbf{0.00059} & 0.10970  \\ 
    ~ & Improvement rate(\%) & +97.68 & +25.38 & +134.30 & +96.67 & -51.04  \\ \hline
    
    \multirow{8}{*}{SZEC} 
    & Buy \& Hold & 0.16288 & 0.01573 & 0.38773 & 0.00066 & 0.18378  \\ 
    ~ & Random Prediction & -0.06758 & -0.00348 & -0.15676 & -0.00027 & 0.11936  \\ 
    ~ & ANN+PCA & 0.06075 & 0.01275 & 0.36238 & 0.00025 & 0.14239  \\ 
    ~ & MKL & 0.12826 & 0.00802 & 0.34007 & 0.00052 & \textbf{0.08520}  \\ 
    ~ & FCN & 0.14902 & 0.00267 & 0.17118 & 0.00060 & 0.09965  \\ 
    ~ & Transformer & 0.12400 & 0.00011 & 0.00425 & 0.00050 & 0.08917  \\
    & TDSE-GA & \textbf{0.32128} & \textbf{0.01850} & \textbf{0.55426} & \textbf{0.00130} & 0.11661  \\ 
    ~ & Improvement rate(\%) & +97.25 & +17.61 & +42.95 & +96.97 & -36.87  \\ \hline
    
    \multirow{8}{*}{GEI} 
    ~ & Buy \& Hold & 0.34745 & 0.02486 & 0.47677 & 0.00141 & 0.22434  \\ 
    ~ & Random Prediction & -0.04239 & 0.00038 & 0.01382 & -0.00017 & 0.15402  \\ 
    ~ & ANN+PCA & 0.22210 & 0.01435 & 0.55245 & 0.00090 & 0.16357  \\ 
    ~ & MKL & 0.12049 & 0.00572 & 0.16746 & 0.00049 & 0.16998  \\ 
    ~ & FCN & 0.25968 & 0.01387 & 0.58775 & 0.00105 & 0.13228  \\ 
    ~ & Transformer & 0.28170 & 0.00569 & 0.13305 & 0.00114 & 0.09032  \\
     & TDSE-GA & \textbf{0.67401} & \textbf{0.03820} & \textbf{0.65072} & \textbf{0.00273} & \textbf{0.08900}  \\ 
    ~ & Improvement rate(\%) & +93.99 & +53.66 & +10.71 & +93.62 & +1.46  \\ \hline
    \end{tabularx}
\end{table}

Based on the results in Table \ref{table:Economic results}, we first find that the trading strategy based on the TDSE-GA model achieves the highest accumulative return on the SSEC (14.49\%), SZEC (32.13\%), and GEI (67.40\%) dataset. These values are 13.6 times, 1.97 times, and 1.94 times that of the Buy \& Hold trading strategy, respectively. In terms of the accumulative return metric, the trading strategy based on the proposed model achieves an improvement of over 90\% compared to the best competing trading strategy. Specifically, the improvement rates are 97.68\% for SSEC, 97.25\% for SZEC, and 93.99\% for GEI. Second, the trading strategy based on the TDSE-GA model provides higher Sharpe ratios compared to other competing trading strategies on the SSEC (0.4059), SZEC (0.5543), and the GEI (0.6507) datasets. This demonstrates that the proposed model can achieve higher returns while taking on a unit of risk. Last, focusing on the maximum drawdown rate, we observe that the optimal trading strategy varies for different SMIs. Specifically, the trading strategy based on the Transformer model performs best on the SSEC (7.26\%), the MKL model on the SZEC (8.52\%), and the TDSE-GA model on the GEI (8.90\%) datasets.

In this study, we use the GA as the optimization method. However, alternative optimization techniques, such as the PSO and FA, are also available. To verify the effectiveness of our chosen algorithm, we compare the GA with these other optimization methods. The comparison criteria include prediction performance and computational time for each model. Detailed results are provided in Section C.4 of the Appendix.

\subsubsection{Dynamic selection process of meta-classifiers}
\begin{figure}
    \centering
    \includegraphics[width=1\linewidth]{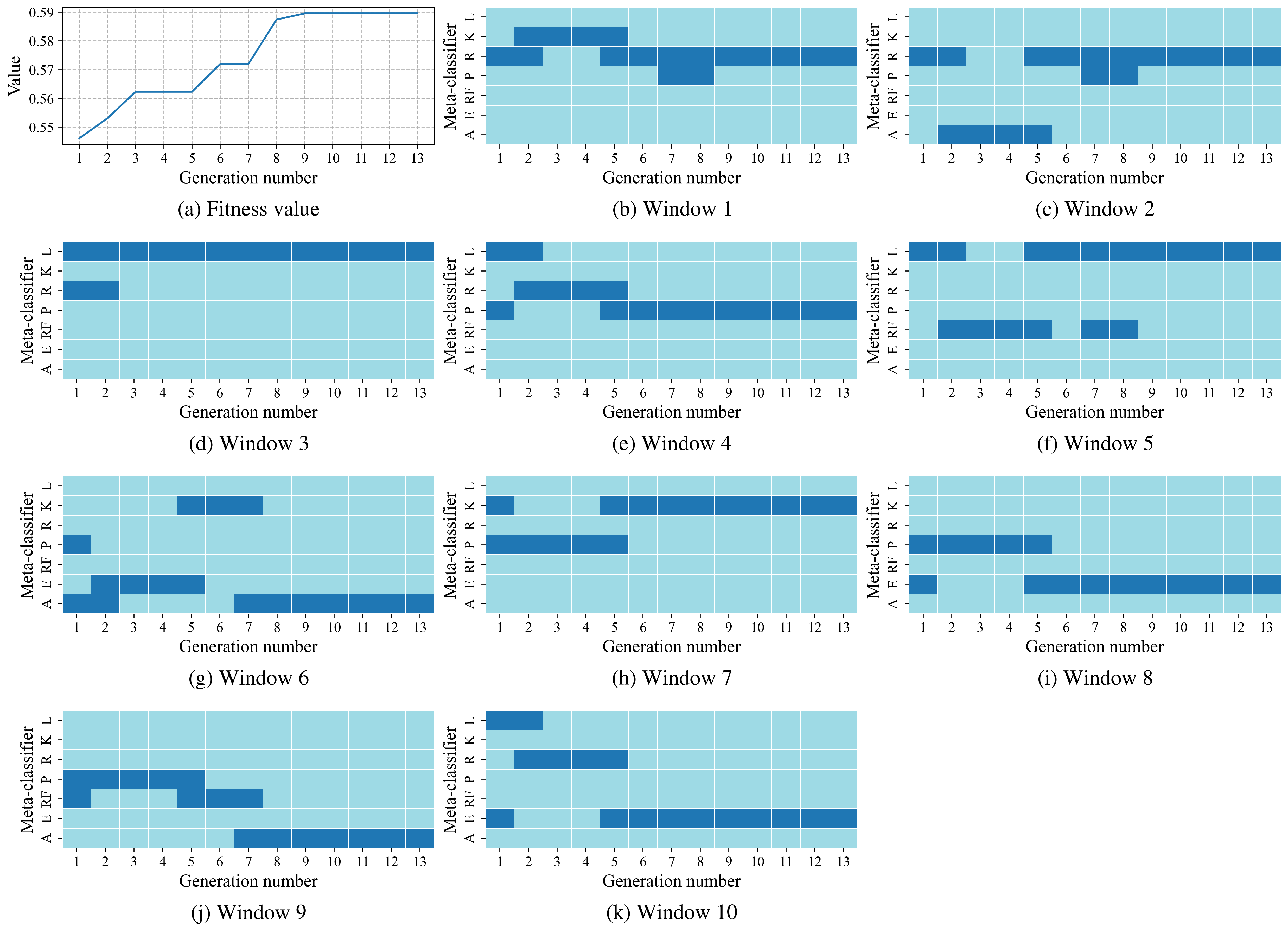}
    \caption{Fitness and selected meta-classifier of the optimal individual over different generations}
    \label{fig:dynamic selection}
\end{figure}

The proposed TDSE-GA model improves prediction performance by progressively determining the optimal classifier at different time periods. Therefore, two critical aspects need to be explored regarding the TDSE-GA model: robustness and dynamic selection process. First, it is important to investigate whether the model's prediction results converge gradually. Second, it is essential to determine which classifiers have significant influence on the model prediction performance at different time periods. To investigate the convergency process of the TDSE-GA model and the dynamic selection process of the meta-classifiers, we carry out the following experiment on the SSEC dataset. First, we extract the fitness function value and the selected meta-classifier of the optimal individual for each generation. Next, we represent these data through a line chart and a heat map, both of which are shown in Fig. \ref{fig:dynamic selection}.

In Fig. \ref{fig:dynamic selection}, the meta-classifiers are represented using abbreviations, corresponding to the following models: L (LR model), K (KNN model), R (RBF-SVM model), P (Poly-SVM model), RF (Random Forest model), E (Extra Trees model), and A (Artificial Neural Network model). Based on the results in Fig. 10, we find the following points. First, in subgraph (a), we notice that the fitness function value consistently increases with the progression of generations, eventually reaching a stable and convergent state. This indicates that the optimization process effectively converges to a satisfactory solution. Second, we observe from subgraph (b) to subgraph (k) that the optimal meta-classifier varies across different time windows, providing evidence of the successful implementation of the dynamic selection design. The adaptive approach ensures that the most suitable meta-classifier is chosen for each specific time period, enhancing the prediction performance. Last, we discover that the LR model. RBF-SVM model, ET model and ANN model are most frequently selected as meta-classifiers (each selected twice). This observation illustrates their superior ability to effectively integrate data from multiple data sources. The dynamic selection verifies their efficiency in improve the overall prediction capabilities of the TDSE-GA model.

\section{Discussion}
\subsection{Theoretical implications}
The theoretical implications of this paper include the understanding of multimodal data for financial prediction, advancements in predictive modeling theories, and implications for behavioral finance theories. In modeling multimodal data, previous studies have highlighted the benefits of information fusion and multimodal learning in enhancing forecasting performance \citep{thakkarFusionStockMarket2021,baoDatadrivenStockForecasting2025}. Building on this foundation, we take into account different data sources, i.e., trend similarities among intraregional SMIs, the industry rotation effect in industry index data, and varying influences from multiple news providers. Compared with the existing research on multimodal data integration \citep{wengPredictingShorttermStock2018,maMultisourceAggregatedClassification2022}, our work incorporates distinct patterns of multimodal data based on investor knowledge. Considering these characteristics, we employ MBCNN, SC-MBCNN, and RNN-ER models to extract inherent features from these data sources. By leveraging the specific attributes of each data type, our model not only improves in predictive accuracy but also provides deeper insights into market dynamics, aiding in more informed decision-making in financial forecasting.

In terms of the implications for the predictive modeling theories, we propose a novel two-stage dynamic stacking ensemble model based on deep learning techniques. This model employs multiple meta-classifiers to dynamically integrate features extracted from each data modality. By accounting for the temporal dynamics between the extracted features and the target SMI movement, our model adapts its selection of the optimal meta-classifier for each time period. This adaptive framework exploits the strengths of different classifiers, harnessing specific patterns identified in each time window to improve the prediction performance. Compared with previous studies \citep{wengPredictingShorttermStock2018,wangCombiningWisdomCrowds2018}, this approach not only demonstrates the practical utility of dynamic ensemble methods in financial forecasting but also contributes to theoretical discussions about the efficacy of ensemble models in capturing complex market dynamics. In addition, we utilize a stage-by-stage optimization method to enhance the generalization capabilities of our proposed model. Initially, we optimize the base classifiers to ensure they are effectively capturing and processing the modal-specific features. Subsequently, we optimize the ensemble component, the meta-classifier, to effectively integrate extracted features from different data sources. This proposed stage-by-stage optimization approach significantly reduces the complexity of the solution space, decreases the risk of overfitting, and improves the model's overall generalization ability. We provide valuable theoretical insights into optimizing complex predictive models, which can be applied broadly across various domains of machine learning and data science.

With respect to the implications for behavioral finance theories, our study delves into how investor experiences and knowledge influence market movement, providing a crucial link to behavioral finance theories. By integrating these behavioral aspects systematically into our predictive models, we not only enhance the models' accuracy but also enrich the understanding of the psychological and cognitive factors that drive financial markets. This approach offers a structured method for quantifying behavioral indicators such as investor sentiment, reaction to news, and decision-making patterns under uncertainty. Furthermore, our research could inspire new theoretical developments by demonstrating the significant role these behavioral elements play in market dynamics.
\subsection{Practical implications}
The practical implications of this study include enhanced market analysis tools that leverage varied data sources for more accurate forecasts, superior predictive performance through advanced multimodal feature extraction, and optimized trading strategies that increase economic value.

First, we find that different data sources exhibit varying predictive abilities, with global SMIs data demonstrating superior performance compared to industry index data and financial news text. This crucial insight offers investors a strategic advantage by identifying the most reliable data sources for market predictions. As a result, this facilitates the advancement of more refined market analysis instruments that can prioritize data sources according to their empirically validated predictive capabilities. These tools could augment the decision-making processes within investment strategy frameworks, fostering more informed financial decisions.

Second, we propose an advanced model designed to extract inherent features from multimodal financial data, enhancing prediction performance. This enhanced capability offers a critical advantage to professional investors by allowing them to refine their analytical models. Integrating this model enables them to achieve more accurate forecasts and improve their risk management practices. Consequently, this leads to more strategic decision-making and potentially greater financial stability and profitability in a volatile market environment.

Last, the trading strategy derived from our proposed model's predictions outperforms alternative strategies, demonstrating both its theoretical robustness and practical economic value. Portfolio managers and algorithmic traders can adopt and integrate this model into their trading systems. By implementing this model, traders could achieve significantly higher returns and enhanced portfolio diversification.

\section{Conclusion and future work}
In this study, we address the SMI movement prediction based on significant factors originated from multi-source financial data. Each financial data source exhibits distinct patterns, which may not be immediately evident from raw financial data but can be discovered by experienced investors. Recognizing and incorporating these data patterns is crucial for enhancing the prediction performance. Hence, we propose the TDSE model, which effectively extracts features from multi-source financial data and improves prediction performance.

Based on the experimental results, we draw the following conclusions. First, different data sources demonstrate varying prediction abilities, with global SMIs data showcasing superior performance compared to industry index data and financial news text. Second, the proposed model provides a remarkable capability in extracting the features of multimodal financial data, resulting in a prediction performance that outperforms the competing methods. Third, among popular optimization algorithms such as PSO, GWO, FA, and HS, the GA algorithm stands out by providing better results in both prediction performance and computational efficiency. Moreover, the trading strategy based on the proposed model surpasses alternative trading strategies, further affirming the economic value of our approach. Last, the proposed model demonstrates swift and stable convergence while effectively identifying the optimal meta-classifier, attesting to its robustness and efficacy. By considering the unique patterns of each financial data source, the proposed model provides a promising solution for enhancing prediction performance of the target SMI movement.

While this study has shown improvements, it is essential to acknowledge its limitations. The following aspects need further attention. (1) This study mainly focuses on the daily frequency, neglecting other frequencies such as minute, hour, or week. To enhance our understanding of multiscale data role in forecasting the target SMI movement, we will explore the significance of these different trading frequencies. (2) This study is centered around the SMI level. However, there is potential value in exploring other levels, including individual stocks and sectors. By expanding our research scope, we can gain comprehensive insights into prediction performance. Based on these studies, we can propose a research framework for portfolio construction that considers prediction results at the individual stock, sector, and SMI level.

\bmhead{Abbreviations} 
\quad\\
2D-CNN \quad 2-Dimension convolutional neural network\\
AUC \quad Area under the ROC curve\\
ANN \quad Artificial neural network\\
BERT \quad Bidirectional encoder representations from transformers\\
BiLSTM \quad Bidirectional long short-term memory network\\
CNN \quad Convolutional neural network\\
FA \quad Firefly algorithm\\
FCN \quad Fully convolutional network\\
FRPCA \quad Fuzzy robust principal component analysis\\
GA \quad Genetic algorithm\\
GEI \quad Growth Enterprise Index\\
GWO \quad Grey wolf optimizer\\  
HS \quad Harmony search algorithm\\
KPCA \quad Kernel principal component analysis\\
LSTM \quad Long short-term memory network\\
MBCNN \quad Multi-branch convolutional neural network\\
MKL \quad Multiple kernel learning\\
PCA \quad Principal component analysis\\
PSO \quad Particle swarm optimization algorithm\\ 
RF \quad Random forest\\
RNN \quad Recurrent neural network\\
RNN-ER \quad Recurrent neural network with event representation\\
SC \quad Spectral clustering\\
SC-MBCNN \quad Spectral clustering-based multi-branch CNN\\
SMI \quad Stock Market Index\\
SSEC \quad Shanghai Securities Composite Index\\
SVM \quad Support vector machine \\
SZEC \quad Shenzhen Component Index\\
TDSE-GA \quad Two-stage dynamic stacking ensemble based on genetic algorithm\\

\bmhead{Acknowledgements}
The study is supported by the National Natural Science Foundation of China (Grant No. 72401029), the China Postdoctoral Science Foundation (Grants No. 2023M740237 and 2024M750254), and the Fundamental Research Funds for the Central Universities (Grants No. 2024CDJSKPT14). 






\begin{appendices}

\section{Description of different data sources}
\label{secA1}

\renewcommand{\thetable}{A.\arabic{table}}

\setlength{\LTcapwidth}{\textwidth} 
{\tiny
\begin{longtable}{ccp{9cm}}

\caption{Description of the global SMIs code}\\
\hline
\textbf{No.} & \textbf{SMIs code} & \textbf{Description} \\
\hline
\endfirsthead
\hline
\textbf{No.} & \textbf{SMIs code} & \textbf{Description} \\
\hline
\endhead
\hline
\endfoot

1 & SSEC & The SSE Composite Index, including all stocks listed on the Shanghai Stock Exchange  \\ 
2 & AXJO & The S\&P/ASX 200 Index, comprising the 200 largest companies listed on the Australian Stock Exchange  \\ 
3 & N225 & The Nikkei 225 Index, including 225 companies listed on the Tokyo Stock Exchange  \\ 
4 & KS11 & The Korea Composite Index, which includes over 780 companies listed on the Korea Stock Exchange  \\
5 & TWII & The Taiwan Weighted Index, which includes all companies listed on the Taiwan Stock Exchange (except for special shares and full delivery shares)  \\
6 & HSI & The Hang Seng Index, which includes blue chip stocks on the Hong Kong Stock Exchange  \\ 
7 & BSESN & The BSE Sensex Index, which consists of 30 companies listed on the Bombay Stock Exchange  \\ 
8 & JKSE & The Jakarta Composite Index, which includes all listed companies on the Indonesia Stock Exchange  \\ 
9 & PSI & The Philippine Manila Index, including the 30 companies with the largest market value on the Philippine Stock Exchange  \\ 
10 & SPX & The S\&P 500 Index, which includes 500 companies listed on the New York Stock Exchange and the Nasdaq Stock Exchange  \\ 
11 & GSPTSE & The S\&P/TSX Index, which includes 250 companies listed on the Toronto Stock Exchange  \\ 
12 & BVSP & The Bovespa Index, which includes the companies listed on the São Paulo Stock Exchange in Brazil  \\ 
13 & COLCAP & The Colombia COLCAP Index, which includes 20 companies listed on the Colombian Stock Exchange  \\ 
14 & MXX & The Mexico BOLSA Index, which includes all companies listed on the Mexican Stock Exchange  \\ 
15 & SPBLPGPT & The Peruvian Lima Index, which includes more than 20 companies listed on the Lima Stock Exchange  \\ 
16 & BEL 20 & The Belgium BEL 20 Index, composed of the 20 most liquid companies listed on the Brussels Stock Exchange  \\
17 & OMXC20 & The OMX Copenhagen 20 Index, which includes the most active companies listed on the Copenhagen Stock Exchange  \\
18 & OMXH25 & The OMX Helsinki 25 Index, including the 25 most actively traded companies on the Helsinki Stock Exchange  \\ 
19 & FCHI & The CAC 40 Index, which includes the 40 largest and the most actively traded companies listed on the Euronext Paris Stock Exchange  \\ 
20 & GDAXI & The DAX 30 Index, which includes the 30 largest and most actively traded companies listed on the Frankfurt Stock Exchange  \\ 
21 & ATG & The Athens Composite Index, comprising 60 companies listed on the Athens Stock Exchange  \\
22 & FTMIB & The FTSE MIB Index, which includes 40 most actively traded companies listed on the Italian Stock Exchange  \\
23 & AEX & The Amsterdam Exchange Index, composed of 25 largest and most actively traded companies listed on the Euronext Amsterdam Stock Exchange  \\
24 & OBX & The Oslo Børs Index, including the 25 most actively traded and liquid stocks listed on the Oslo Stock Exchange  \\ 
25 & WIG30 & The Warszawski Indeks Gieldowy Index, composed of the 30 largest and most liquid companies listed on the Warsaw Stock Exchange  \\ 
26 & PSI20 & The Portuguese Stock Index, which includes the 20 largest and most liquid companies listed on the Euronext Lisbon exchange  \\
27 & IMOEX & The MOEX Russia Index, including the largest and most liquid listed companies on the Moscow Exchange  \\ 
28 & IBEX 35 & The IBerian IndEX Index, composed of the 35 most liquid and actively traded stocks listed on the Madrid Stock Exchange  \\ 
29 & BIST 100 & The Borsa Istanbul 100 Index, which includes the 100 most liquid and actively traded stocks listed on the Istanbul Stock Exchange  \\
30 & FTSE & The Financial Times Stock Exchange 100 Index, comprising the 100 largest companies listed on the London Stock Exchange  \\ \hline
\end{longtable}
}

\setlength{\LTcapwidth}{\textwidth} 
{\tiny
\begin{longtable}{p{5cm}cp{6cm}}
\caption{Description of the industry index data}\\
\hline
\textbf{Industry name} & \textbf{No.} & \textbf{Sub-industry name} \\
\hline
\endfirsthead
\hline
\textbf{Industry name} & \textbf{No.} & \textbf{Sub-industry name} \\
\hline
\endhead
\hline
\endfoot
Agriculture, forestry, animal husbandry and fishery & 1-1 & Agriculture  \\ 
~ & 1-2 & Forestry  \\ 
~ & 1-3 & Animal Husbandry  \\ 
~ & 1-4 & Fishery  \\ 
~ & 1-5 & Agriculture, Forestry, Animal Husbandry and Fishery Services  \\ \hline

Mining & 2-1 & Coal mining and washing industry  \\ 
        ~ & 2-2 & Oil and gas mining industry  \\ 
        ~ & 2-3 & Ferrous metal mining industry  \\ 
        ~ & 2-4 & Non-ferrous metal mining industry  \\ 
        ~ & 2-5 & Non-metallic mining industry  \\ 
        ~ & 2-6 & Mining auxiliary activities  \\ \hline

Manufacturing & 3-1 & Agricultural and food processing industry  \\ 
        ~ & 3-2 & Food manufacturing industry  \\ 
        ~ & 3-3 & Wine, beverage and refined tea manufacture industry  \\ 
        ~ & 3-4 & Textile industry  \\ 
        ~ & 3-5 & Textile clothing, apparel industry  \\ 
        ~ & 3-6 & Leather, fur, feathers and their products and shoe industry  \\ 
        ~ & 3-7 & Wood processing and wood, bamboo, rattan, palm, grass products industry  \\ 
        ~ & 3-8 & Furniture manufacture industry  \\ 
        ~ & 3-9 & Paper and paper products industry  \\ 
        ~ & 3-10 & Printing and recording media reproduction industry  \\ 
    ~ & 3-11 & Education, sports and entertainment goods manufacturing industry  \\
        ~ & 3-12 & Petroleum processing, coking and nuclear fuel processing industry  \\ 
        ~ & 3-13 & Chemical raw materials and chemical products manufacturing industry  \\ 
        ~ & 3-14 & Pharmaceutical manufacturing industry  \\ 
        ~ & 3-15 & Chemical fiber manufacturing industry  \\ 
        ~ & 3-16 & Rubber and plastic products industry  \\ 
        ~ & 3-17 & Non-metallic mineral products industry  \\ 
        ~ & 3-18 & Ferrous metal smelting and rolling processing industry  \\
        ~ & 3-19 & Non-ferrous metal smelting and rolling processing industry  \\
        ~ & 3-20 & Metal products industry  \\ 
         ~ & 3-21 & General equipment manufacturing industry  \\ 
        ~ & 3-22 & Special equipment manufacturing industry  \\ 
        ~ & 3-23 & Automobile manufacturing industry  \\ 
        ~ & 3-24 & Railroad, ship, aerospace and other transportation equipment manufacturing industry  \\ 
        ~ & 3-25 & Electrical machinery and equipment manufacturing industry  \\ 
        ~ & 3-26 & Computer, communication and other electronic equipment manufacturing industry  \\ 
        ~ & 3-27 & Instrument manufacturing  \\ 
        ~ & 3-28 & Other manufacturing industries  \\ 
        ~ & 3-29 & Comprehensive utilization of waste resources industry  \\ \hline

Electricity, heat, gas and water production and supply industry & 4-1 & Electricity, heat production and supply industry \\  
    ~ & 4-2 & Gas production and supply industry  \\ 
    ~ & 4-3 & Water production and supply industry  \\ \hline
Construction & 5-1 & Housing construction industry  \\ 
~ & 5-2 & Civil engineering construction industry  \\ 
~ & 5-3 & Construction and installation industry  \\
~ & 5-4 & Building decoration and other construction industry  \\ \hline

Wholesale and retail trade & 6-1 & Wholesale  \\ 
        ~ & 6-2 & Retail trade  \\ \hline
Transportation, storage and postal services & 7-1 & Rail transportation  \\ 
        ~ & 7-2 & Road transportation  \\ 
        ~ & 7-3 & Water transportation  \\ 
        ~ & 7-4 & Air transportation  \\ 
        ~ & 7-5 & Stevedoring and other transport agency industry  \\ 
        ~ & 7-6 & Warehousing  \\ 
        ~ & 7-7 & Postal industry  \\ \hline

Accommodation and Catering & 8-1 & Accommodation  \\
        ~ & 8-2 & Catering  \\ \hline
Information transmission, software and information technology services & 9-1 & Telecommunications, radio and television and satellite transmission services  \\ 
        ~ & 9-2 & Internet and related services  \\
        ~ & 9-3 & Software and Information Technology Services  \\ \hline
Finance & 10-1 & Monetary and Financial Services  \\
        ~ & 10-2 & Capital Market Services  \\ 
        ~ & 10-3 & Insurance  \\ 
        ~ & 10-4 & Other Financial Services  \\ \hline
Leasing and Business Services & 11-1 & Leasing  \\ 
        ~ & 11-2 & Business Services  \\ \hline
Scientific Research and Technology Services & 12-1 & Research and Experimental Development  \\ 
        ~ & 12-2 & Professional and Technical Services  \\ 
        ~ & 12-3 & Science and Technology Promotion and Application Services  \\ \hline
Water, Environment and Public Facilities Management & 13-1 & Ecological protection and environmental management  \\ 
        ~ & 13-2 & Public Facilities Management  \\ \hline
Health and Social Work & 14-1 & Health  \\ 
    ~ & 14-2 & Social Work  \\ \hline
Culture, Sports and Entertainment & 15-1 & Journalism and publishing  \\
    ~ & 15-2 & Radio, television, film and video recording production  \\ 
    ~ & 15-3 & Culture and Arts  \\ 
    ~ & 15-4 & Sports  \\ \hline
Real Estate industry & ~ & Real Estate  \\ 
Education & ~ & Education  \\ 
Residential services, repair and & ~ & Motor vehicles, electronic products and daily use  \\ 
General industry & ~ & General industry  \\ \hline
\end{longtable}
}

\begin{table}[!ht]
\centering
\caption{Description of the financial news text}
\footnotesize 
    \begin{tabular}{l p{10cm}}
    \hline
    News provider & Input description  \\ \hline
    Sina Finance & Positive index, Negative index, Neutral index, News number of Sina Finance; Historical data of SMI (Open price, high price, low price, close price, trading volume, daily return); The lag period is $t-1,t-2,\ldots,t-L$.  \\ 
    Wall Street CN & Positive index, Negative index, Neutral index, News number of Wall Street CN; Historical data of SMI (Open price, high price, low price, close price, trading volume, daily return); The lag period is $t-1,t-2,\ldots,t-L$.  \\ 
    Straight flush & Positive index, Negative index, Neutral index, News number of Straight flush; Historical data of SMI (Open price, high price, low price, close price, trading volume, daily return); The lag period is $t-1,t-2,\ldots,t-L$.  \\ 
    East money & Positive index, Negative index, Neutral index, News number of East Money; Historical data of SMI (Open price, high price, low price, close price, trading volume, daily return); The lag period is $t-1,t-2,\ldots,t-L$.  \\ 
    Yun CaiJing & Positive index, Negative index, Neutral index, News number of Yun CaiJing; Historical data of SMI (Open price, high price, low price, close price, trading volume, daily return); The lag period is t-1,$t-1,t-2,\ldots,t-L$.  \\ \hline
    \end{tabular}
\end{table}

\section{Prediction models}
\renewcommand{\thetable}{B.\arabic{table}}
\setcounter{equation}{0}  
\renewcommand{\theequation}{B.\arabic{equation}}  

\subsection*{B.1 Logistic regression model}
Given the input vectors $\boldsymbol{x}=(x_{1}, x_{2}, \ldots, x_{6})$, the linear model is shown in the follow equations: 
\begin{equation}
    z=w_{1} x_{1}+w_{2} x_{2}+\ldots+w_{d} x_{d}+b=\boldsymbol{w}^{T} \boldsymbol{x}+b,
\end{equation}
where $\boldsymbol{w}=(w_{1}, w_{2}, \ldots, w_{d})$ is the weight vector, $d=6$ and the $b$ is the bias. The research problem of this study is a binary problem, and the output of the proposed model is the target SMI movement denoted as $y \in\{0,1\}$. Based on the Logistic function, $y$ is represented by the following equation:
\begin{equation}
    y=\frac{1}{1+e^{-z}}=\frac{1}{1+e^{-(\boldsymbol{w}^{T} \boldsymbol{x}+b)}}.
\end{equation}

Based on the logarithmic change, the Eq. (B.2) is transformed into the following equation: 
\begin{equation}
    \ln \frac{y}{1-y}=\boldsymbol{w}^{T} \boldsymbol{x}+b.
\end{equation}

If the $y$ is viewed as the posterior probability estimation $p(y=1 \mid \boldsymbol{x})$, the Eq. (B.3) is transformed into the following equation: 
\begin{equation}
    \ln \frac{p(y=1 \mid \boldsymbol{x})}{p(y=0 \mid \boldsymbol{x})}=\boldsymbol{w}^{T} \boldsymbol{x}+b.
    \label{eq:logistic prob}
\end{equation}

Based on the Eq. (\ref{eq:logistic prob}), the parameters $\boldsymbol{w}$ and $b$ are determined by the maximum likelihood method. Finally, based on the parameters $\boldsymbol{w}$ and $b$ , we can obtain the prediction results by the LR model.

\subsection*{B.2 K-nearest neighbor model}
Given the train set $\{(\boldsymbol{x}_{1}, y_{1}),(\boldsymbol{x}_{2}, y_{2}), \ldots,(\boldsymbol{x}_{n}, y_{n})\}$ where $\boldsymbol{x}_i$ is the probability input vector of the $i\text{-}th$ sample, $y_i$ is the class of the $i\text{-}th$ sample. First, the KNN model finds $k$ nearest neighbors to the point $\boldsymbol{x}_{i}$ in the training dataset. The neighborhood $N_{k}(\boldsymbol{x}_{i})$ is made up of the above $k$ nearest neighbors where the distance is calculated by the Euclidean distance:
\begin{equation}
    L(\boldsymbol{x}_{i}, \boldsymbol{x}_{j})=\sqrt{\sum_{l=1}^{d}(\boldsymbol{x}_{i}^{l}-\boldsymbol{x}_{j}^{l})^{2}}.
\end{equation}

Based on the majority rule in the neighborhood $N_{k}(\boldsymbol{x}_{i})$, the unknown class is determined by the majority of the $k$ nearest neighbors to the point $\boldsymbol{x}_i$:
\begin{equation}
    y_{i}=\arg \max _{c_{j}} \sum_{x_{a} \in N_{k}(x_{i})} I(y_{a}=c_{j}),
\end{equation}
where $I$ is the indicator function. If $y_{a}=c_{j}$, $I=1$, else $I=0$.

\subsection*{B.3 Support vector machine model}
Given the train set $\{(\boldsymbol{x}_{1}, y_{1}),(\boldsymbol{x}_{2}, y_{2}), \ldots,(\boldsymbol{x}_{n}, y_{n})\}$, suppose there is not exist a hyperplane that correctly divides the two class samples in the original feature space.  And then, suppose the $\phi(\boldsymbol{x})$ is the mapped feature vector of $\boldsymbol{x}$ , the hyperplane of the feature space is shown in the following equation:
\begin{equation}
    f(\boldsymbol{x})=\boldsymbol{w}^{T} \phi(\boldsymbol{x})+b,
    \label{eq:hyperplane}
\end{equation}
where $\boldsymbol{w}$ and $b$ are the weight vector and the bias. To determine the separating hyperplane with the maximum margin, Eq. (\ref{eq:hyperplane}) needs to be satisfied by the following equation: 
\begin{equation}
\begin{aligned}
    \min_{\boldsymbol{w}, b} \quad & \frac{1}{2}\|\boldsymbol{w}\|^{2} \\
    \text{s.t.} \quad & y_{i}(\boldsymbol{w}^{T} \phi(\boldsymbol{x}_{i}) + b) \geq 1, \quad i=1,2,\ldots,n
\end{aligned}
\end{equation}
where the support vectors are these samples ($y_{i}(\boldsymbol{w}^{T} \phi(\boldsymbol{x}_{i})+b)=1$). Based on the Lagrange multiplier method and the soft margin, the dual problem of the Eq. (\ref{eq:SVM-dual}) is shown in the following equation: 
\begin{equation}
\begin{aligned}
\max _{\alpha} \sum_{i=1}^{n} \alpha_{i}-\frac{1}{2} \sum_{i=1}^{n} \sum_{j=1}^{n} \alpha_{i} \alpha_{j} y_{i} y_{j} \phi(\boldsymbol{x}_{i})^{T} \phi(\boldsymbol{x}_{j})\\
s.t. \sum_{i=1}^{n} \alpha_{i} y_{i}=0 \\
0 \leq \alpha_{i} \leq C, i=1,2, \ldots, \mathrm{n}
\label{eq:SVM-dual}
\end{aligned}
\end{equation}
where the $\alpha_{i}$ is the Lagrange multiplier, $C$ is the regularization constant. Because the number of dimension in the feature space might be high, which increases the difficulty to calculate, the SVM model employs the kernel function to calculate the $\phi(\boldsymbol{x}_{i})^{T} \phi(\boldsymbol{x}_{j})$:
\begin{equation}
    k(x_{i}, x_{j})=\phi(\boldsymbol{x}_{i})^{T} \phi(\boldsymbol{x}_{j})
\end{equation}

The kernel function generally includes the linear kernel function, polynomial kernel function, radial basis kernel function, sigmoid kernel function, and Laplacian kernel function. If the solution of Eq. (\ref{eq:SVM-dual}) satisfies the Karush-Kuhn-Tucker condition, the solution of the original problem and the dual problem can be obtained. Based on the trained parameters $\boldsymbol{w}$ and $b$, we can obtain the prediction results of the SVM model.

\subsection*{B.4 Random forests and extratrees model}
The random forests (RF) model and the extratrees (ET) model which are both based on the decision tree model include multiple decision tree models. The prediction result of the above models is determined by the mode of output type of multiple decision trees. A decision tree mainly includes a root node, many internal nodes and leaf nodes. where the leaf node is the decision result, other type of nodes are the attribute test. The attributes of a decision tree are divided recursively to construct a test sequence from the root node to the leaf node. Because the decision tree model uses all samples and all features to train the model, the overfitting risk will be increased. To decrease the overfitting risk, the RF model and the ET model randomly generate multiple decision trees. We describe the details of the RF model and the ET model as shown as follows.

For the RF model, given a dataset with $n$ samples where each sample has $K$ features, each decision tree randomly select $m$ train samples from the dataset with replacement ($m<n$) and uses randomly $k$ features ($k<K$). Based on the selected samples and features, a decision tree can be constructed. And then, the stochastic process is repeated to build multiple decision trees. The diversity of generated decision trees is guaranteed by the sample randomness and the feature randomness.

The difference between the ET model and the RF model includes the following two points: (1) The RF model selects the samples with the replacement way, while the ET model uses all samples. (2) Based on the selected features, the RF model recursively choose the best splitting feature to build a decision tree, while the ET model randomly select the feature to build a decision tree.

\subsection*{B.5 Artificial neural network model}
Given the train dataset $\{(\boldsymbol{x}_{1}, y_{1}),(\boldsymbol{x}_{2}, y_{2}), \ldots,(\boldsymbol{x}_{n}, y_{n})\}$, the first layer of the ANN model is the input layer which includes six neurons $\boldsymbol{x}=(x_{1}, x_{2}, \ldots, x_{6})$ There are three hidden layers in the designed ANN model. Based on the input layer, the output of the $j\text{-}th$ neuron of the first hidden layer is calculated by the following formula:
\begin{equation}
    o_{j}^{(1)}=\theta(\sum_{i=1}^{d} \boldsymbol{w}_{i, j}^{(1)} \boldsymbol{x}_{i}+b_{j})
\end{equation}
where the $\theta$ is the activation function, $\boldsymbol{w}_{i, j}$ is the connection weight of the $i\text{-}th$ neuron in the first hidden layer and the $j\text{-}th$ neuron in the second hidden layer, $b_j$ is the bias. The output of neurons in the second hidden layer and in the third hidden layer are also calculated in the above formula. The last layer is the output layer which includes two neurons. In the last layer, the softmax function is employed as the activation function, and the probability of the $i\text{-}th$ type is calculated by the following formula: 
\begin{equation}
    p_{i}=\frac{e^{z_{i}}}{\sum_{j=1}^{n} e^{z_{j}}}
\end{equation}
where the $n$ is the number of types, $z_j$ is the output value of the $j\text{-}th$ neuron. Based on the output layer, the type with the maximum support probability is used as the prediction result, i.e., the predicted SMI movement.

\section{Detailed experimental results}
\renewcommand{\thetable}{C.\arabic{table}}
\setcounter{table}{0}
\renewcommand{\thefigure}{C.\arabic{figure}}
\setcounter{figure}{0}

\subsection*{C.1. Data description}
Table \ref{table:statistical results on the sentiment} presents the statistical results on the sentiment indexes and news volume of different financial news providers.

\begin{table}[!ht]
\centering
\setlength{\belowcaptionskip}{10pt}
\caption{The statistical results on the sentiment indexes and news volume}
\label{table:statistical results on the sentiment}
{\tiny
\begin{tabular}
    {c c c c c c c c c c c c}
    \hline
        Media measure & Mean & Std & 25\% & Median & 75\% & ~ & Mean & Std & 25\% & Median & 75\%  \\
        
    \hline
        ~ & \multicolumn{5}{c}{News provider 1: Sina Finance} & ~ & \multicolumn{5}{c}{News provider 2: Wall Street CN}\\
        
    
        Positive index & 0.502 & 0.070 & 0.445 & 0.505 & 0.551 & ~ & 0.519 & 0.060 & 0.481 & 0.518 & 0.561\\ 
        Negative index & 0.220 & 0.049 & 0.185 & 0.216 & 0.248 & ~ & 0.209 & 0.040 & 0.180 & 0.205 & 0.230\\ 
        Neutral index & 0.279 & 0.041 & 0.250 & 0.281 & 0.307 & ~ & 0.273 & 0.041 & 0.244 & 0.271 & 0.299\\ 
        Number of news & 546.6 & 330.5 & 220.5 & 556.0 & 901.5 & ~ & 351.3 & 213.8 & 203.0 & 322.0 & 418.5\\
        
    
        ~ & \multicolumn{5}{c}{News provider 3: Straight flush} & ~ & \multicolumn{5}{c}{News provider 4: East money}\\ 
        
    
        Positive index & 0.534 & 0.064 & 0.497 & 0.538 & 0.573 & ~ & 0.540 & 0.060 & 0.499 & 0.541 & 0.585\\ 
        Negative index & 0.203 & 0.052 & 0.165 & 0.198 & 0.232 & ~ & 0.222 & 0.048 & 0.187 & 0.217 & 0.253\\ 
        Neutral index & 0.263 & 0.041 & 0.241 & 0.263 & 0.288 & ~ & 0.238 & 0.045 & 0.206 & 0.232 & 0.266\\ 
        Number of news & 294.9 & 136.9 & 164.0 & 326.0 & 405.5 & ~ & 331.4 & 103.8 & 297.5 & 349.0 & 385.0\\ 
        
    
        ~ & \multicolumn{5}{c}{News provider 5: Yun CaiJing} & ~ & \multicolumn{5}{c}{Market news}\\ 
        
    
        Positive index & 0.567 & 0.060 & 0.533 & 0.571 & 0.601 & ~ & 0.525 & 0.055 & 0.489 & 0.524 & 0.562\\ 
        Negative index & 0.175 & 0.041 & 0.147 & 0.169 & 0.197 & ~ & 0.208 & 0.040 & 0.180 & 0.205 & 0.230\\ 
        Neutral index & 0.258 & 0.039 & 0.238 & 0.259 & 0.280  & ~ & 0.267 & 0.035 & 0.244 & 0.268 & 0.291\\
        Number of news & 348.8 & 135.5 & 285.0 & 370.0 & 440.5 & ~ & 1404 & 588.1 & 977.5 & 1481.0 & 1835\\ 
        
    \hline
\end{tabular}
}
\end{table}

Based on the results of Table \ref{table:statistical results on the sentiment}, we find that the average positive value of the five news providers is higher than the negative and neutral indexes, which illustrates that these Chinese news providers are inclined to publish news articles with an enthusiastic tone. In contrast, previous studies have shown that English news media tends to publish more pessimistic news\footnote{Gao, R., Cui, S., Xiao, H., Fan, W., Zhang, H., \& Wang, Y. (2022). Integrating the sentiments of multiple news providers for stock market index movement prediction: A deep learning approach based on evidential reasoning rule. Information Sciences, 615, 529–556.}. Furthermore, our analysis of Table \ref{table:statistical results on the sentiment} reveals that Sina Finance publishes a greater number of daily news articles compared to the other news providers.

\subsection*{C.2. Prediction results of different data sources}

Specifically, we employ the above data sources as inputs into a range of forecasting models, including ANN, SVM, RF, CNN, RNN, LSTM, GRU, and 2D-CNN. The prediction results of different data sources for the SSEC, SZEC, and GEI are shown in Table \ref{table:prediction results of different data sources} where "Market" means that the input of model is the historical market data, "Global" denotes that the input of model is the global SMIs data, "Industry" indicates the industry index data as input, and "News" refers to the sentiment index of different news providers as input. In Table \ref{table:prediction results of different data sources}, the underlined number in each row represents the best value for that specific row. The "Mean" metric represents the average value of all prediction model in each row. The number in boldface within the "Mean" column indicates the maximum prediction result of different data sources for an evaluation metric. Additionally, the "Rank" metric represents the sorting result of the "Mean" metric with respect to different data sources.

\setlength{\LTcapwidth}{\textwidth}  
{\tiny
\begin{longtable}{cl cccc cccc cc}
\caption{The prediction results of different data sources for the SSEC, SZEC, and GEI} 
\label{table:prediction results of different data sources} \\

\hline
Metric & Input & ANN & SVM & RF & CNN & RNN & LSTM & GRU & 2D-CNN & Mean & Rank\\
\hline
\endfirsthead

\hline
Metric & Input & ANN & SVM & RF & CNN & RNN & LSTM & GRU & 2D-CNN & Mean & Rank\\
\hline
\endhead


\hline
\endlastfoot

\multicolumn{12}{c}{Target SMI: SSEC} \\ \hline
\multirow{4}{*}{Accuracy}
& Market & 0.496 & 0.525 & 0.511 & 0.513 & \underline{0.527} & 0.498 & 0.521 & 0.486 & 0.510 & 3  \\ 
~ & Global & 0.525 & \underline{0.568} & 0.521 & 0.527 & 0.502 & 0.536 & 0.519 & 0.519 & \textbf{0.527} & 1  \\ 
~ & Industry & 0.498 & 0.458 & 0.491 & 0.501 & 0.494 & \underline{0.504} & 0.497 & 0.483 & 0.491 & 4  \\ 
~ & News & 0.497 & \underline{0.531} & 0.526 & 0.512 & 0.519 & 0.494 & 0.501 & 0.511 & 0.511 & 2  \\ \hline

\multirow{4}{*}{Recall} 
& Market & 0.451 & 0.518 & 0.426 & 0.578 & 0.460 & 0.537 & 0.449 & \underline{0.581} & 0.500 & 3  \\ 
~ & Global & 0.546 & 0.605 & \underline{0.619} & 0.605 & 0.514 & 0.581 & 0.583 & 0.607 & \textbf{0.582} & 1  \\ 
~ & Industry & 0.554 & 0.469 & 0.556 & 0.543 & 0.517 & 0.589 & \underline{0.592} & 0.539 & 0.545 & 2  \\ 
~ & News & 0.433 & 0.446 & 0.407 & 0.520 & 0.457 & 0.463 & 0.413 & \underline{0.531} & 0.459 & 4  \\ \hline

\multirow{4}{*}{Precision} 
& Market & 0.486 & 0.528 & 0.416 & 0.498 & 0.476 & 0.516 & \underline{0.562} & 0.454 & 0.492 & 3  \\ 
~ & Global & 0.508 & \underline{0.554} & 0.521 & 0.520 & 0.494 & 0.525 & 0.515 & 0.518 & \textbf{0.519} & 1  \\ 
~ & Industry & 0.496 & 0.454 & 0.489 & 0.499 & 0.496 & \underline{0.500} & 0.493 & 0.480 & 0.488 & 4  \\ 
~ & News & 0.498 & 0.544 & \underline{0.549} & 0.505 & 0.518 & 0.489 & 0.484 & 0.544 & 0.517 & 2  \\ \hline

\multirow{4}{*}{\textit{F}-measure} 
& Market & 0.449 & 0.496 & 0.399 & \underline{0.513} & 0.426 & 0.459 & 0.411 & 0.468 & 0.452 & 3.5  \\ 
~ & Global & 0.521 & \underline{0.570} & 0.551 & 0.543 & 0.494 & 0.540 & 0.533 & 0.543 & \textbf{0.537} & 1  \\ 
~ & Industry & 0.511 & 0.457 & 0.507 & 0.512 & 0.499 & \underline{0.533} & 0.532 & 0.496 & 0.506 & 2  \\
~ & News & 0.433 & 0.463 & 0.420 & 0.481 & 0.458 & 0.451 & 0.420 & \underline{0.491} & 0.452 & 3.5  \\ \hline

\multirow{4}{*}{AUC} 
& Market & 0.518 & 0.509 & 0.541 & 0.544 & 0.559 & \underline{0.577} & 0.567 & 0.569 & 0.548 & 2  \\ 
~ & Global & 0.491 & \underline{0.597} & 0.542 & 0.546 & 0.540 & 0.561 & 0.545 & 0.567 & \textbf{0.549} & 1  \\ 
~ & Industry & 0.516 & 0.459 & \underline{0.519} & 0.507 & 0.473 & 0.509 & 0.485 & 0.471 & 0.492 & 4  \\ 
~ & News & 0.495 & 0.505 & \underline{0.551} & 0.506 & 0.483 & 0.507 & 0.466 & 0.527 & 0.505 & 3  \\ \hline

\multicolumn{12}{c}{Target SMI: SZEC} \\ \hline
\multirow{4}{*}{Accuracy} 
& Market & 0.480 & \underline{0.521} & 0.493 & 0.488 & 0.488 & 0.495 & 0.512 & 0.487 & 0.496 & 3.5  \\ 
~ & Global & 0.529 & \underline{0.562} & 0.557 & 0.531 & 0.533 & 0.542 & 0.551 & 0.532 & \textbf{0.542} & 1  \\ 
~ & Industry & 0.483 & 0.489 & 0.496 & \underline{0.516} & 0.504 & 0.491 & 0.510 & 0.511 & 0.500 & 2  \\ 
~ & News & 0.501 & 0.481 & 0.495 & 0.500 & 0.501 & 0.480 & 0.490 & \underline{0.519} & 0.496 & 3.5  \\ \hline

\multirow{4}{*}{Recall} & Market & 0.382 & 0.399 & 0.300 & 0.462 & 0.436 & \underline{0.482} & 0.475 & 0.454 & 0.424 & 3  \\ 
~ & Global & 0.596 & 0.608 & \underline{0.611} & 0.605 & 0.541 & 0.575 & 0.581 & 0.564 & \textbf{0.585} & 1  \\ 
~ & Industry & 0.469 & 0.479 & 0.499 & \underline{0.558} & 0.506 & 0.517 & 0.534 & 0.533 & 0.512 & 2  \\ 
~ & News & 0.416 & 0.313 & 0.342 & 0.435 & 0.472 & 0.404 & 0.455 & \underline{0.528} & 0.421 & 4  \\ \hline

\multirow{4}{*}{Precision} & Market & 0.472 & 0.456 & 0.499 & 0.525 & 0.511 & 0.472 & 0.565 & \underline{0.595} & 0.512 & 3  \\ 
~ & Global & 0.544 & \underline{0.578} & 0.571 & 0.534 & 0.550 & 0.550 & 0.565 & 0.545 & \textbf{0.555} & 1  \\ 
~ & Industry & 0.512 & 0.516 & 0.515 & 0.532 & 0.525 & 0.518 & \underline{0.538} & \underline{0.538} & 0.524 & 2  \\ 
~ & News & 0.510 & 0.460 & 0.502 & 0.483 & 0.519 & 0.493 & 0.490 & \underline{0.521} & 0.497 & 4  \\ \hline

\multirow{4}{*}{\textit{F}-measure} & Market & 0.391 & 0.405 & 0.334 & 0.418 & 0.399 & \underline{0.431} & 0.430 & 0.396 & 0.400 & 4  \\ 
~ & Global & 0.563 & 0.581 & \underline{0.581} & 0.557 & 0.540 & 0.555 & 0.564 & 0.543 & \textbf{0.561} & 1  \\ 
~ & Industry & 0.483 & 0.490 & 0.501 & \underline{0.540} & 0.508 & 0.507 & 0.525 & 0.527 & 0.510 & 2  \\ 
~ & News & 0.434 & 0.355 & 0.358 & 0.407 & 0.476 & 0.417 & 0.445 & \underline{0.499} & 0.424 & 3  \\ \hline

\multirow{4}{*}{AUC} & Market & 0.487 & 0.533 & 0.516 & 0.518 & \underline{0.548} & 0.507 & 0.529 & 0.509 & 0.518 & 2  \\ 
~ & Global & 0.546 & \underline{0.590} & 0.586 & 0.541 & 0.556 & 0.552 & 0.558 & 0.556 & \textbf{0.561} & 1  \\ 
~ & Industry & 0.498 & 0.487 & 0.487 & 0.481 & 0.492 & 0.463 & 0.486 & \underline{0.499} & 0.487 & 3  \\ 
~ & News & 0.494 & \underline{0.502} & 0.489 & 0.436 & 0.435 & 0.466 & 0.452 & 0.478 & 0.469 & 4  \\ \hline

\multicolumn{12}{c}{Target SMI: GEI} \\ \hline
\multirow{4}{*}{Accuracy} 
& Market & 0.544 & 0.493 & 0.561 & 0.533 & 0.561 & 0.558 & \underline{0.583} & 0.507 & 0.542 & 2  \\ 
~ & Global & 0.550 & \underline{0.578} & 0.561 & 0.572 & 0.517 & 0.570 & 0.559 & 0.541 & \textbf{0.556} & 1  \\ 
~ & Industry & 0.524 & 0.519 & 0.509 & 0.522 & 0.553 & 0.539 & 0.525 & \underline{0.559} & 0.531 & 3  \\ 
~ & News & 0.517 & 0.496 & \underline{0.523} & 0.497 & 0.515 & 0.513 & 0.511 & 0.509 & 0.510 & 4  \\ \hline

\multirow{4}{*}{Recall} 
& Market & 0.521 & 0.336 & 0.480 & 0.620 & 0.613 & \underline{0.725} & 0.650 & 0.560 & 0.563 & 3.5  \\ 
~ & Global & 0.540 & 0.646 & 0.679 & 0.662 & 0.562 & 0.659 & 0.624 & \underline{0.685} & \textbf{0.632} & 1  \\ 
~ & Industry & 0.564 & 0.558 & \underline{0.622} & 0.597 & 0.547 & 0.581 & 0.590 & 0.574 & 0.579 & 2  \\ 
~ & News & 0.534 & 0.491 & 0.468 & 0.599 & 0.591 & 0.611 & 0.597 & \underline{0.611} & 0.563 & 3.5  \\ \hline

\multirow{4}{*}{Precision} 
& Market & 0.615 & 0.477 & 0.583 & 0.600 & 0.613 & 0.603 & \underline{0.652} & 0.502 & 0.581 & 3  \\ 
~ & Global & 0.601 & \underline{0.620} & 0.596 & 0.611 & 0.573 & 0.611 & 0.608 & 0.577 & \textbf{0.600} & 1  \\ 
~ & Industry & 0.580 & 0.581 & 0.560 & 0.569 & \underline{0.618} & 0.599 & 0.573 & 0.610 & 0.586 & 2  \\ 
~ & News & 0.525 & 0.541 & \underline{0.687} & 0.526 & 0.551 & 0.562 & 0.562 & 0.566 & 0.565 & 4  \\ \hline

\multirow{4}{*}{\textit{F}-measure} 
& Market & 0.517 & 0.376 & 0.491 & 0.548 & 0.567 & \underline{0.615} & 0.591 & 0.505 & 0.526 & 3.5  \\ 
~ & Global & 0.563 & 0.620 & 0.622 & \underline{0.625} & 0.558 & 0.623 & 0.606 & 0.616 & \textbf{0.604} & 1  \\ 
~ & Industry & 0.561 & 0.558 & 0.575 & 0.572 & 0.571 & 0.575 & 0.572 & \underline{0.585} & 0.571 & 2  \\ 
~ & News & 0.509 & 0.500 & 0.460 & 0.544 & 0.552 & \underline{0.560} & 0.547 & 0.535 & 0.526 & 3.5  \\ \hline

\multirow{4}{*}{AUC} 
& Market & 0.529 & 0.492 & 0.537 & 0.545 & 0.591 & 0.582 & \underline{0.596} & 0.534 & 0.551 & 2  \\ 
~ & Global & 0.569 & \underline{0.618} & 0.589 & 0.556 & 0.535 & 0.592 & 0.585 & 0.522 & \textbf{0.571} & 1  \\ 
~ & Industry & 0.498 & 0.529 & 0.514 & 0.529 & 0.562 & \underline{0.571} & 0.548 & 0.551 & 0.538 & 3  \\ 
~ & News & 0.510 & 0.483 & \underline{0.514} & 0.486 & 0.505 & 0.502 & 0.480 & 0.494 & 0.497 & 4  \\ \hline
\end{longtable}
}

According to results in Table \ref{table:prediction results of different data sources}, our analysis yields two crucial insights. First, we investigate the impact of integrating supplementary data sources alongside historical market data on the prediction performance. We find the influence of these diverse data sources varies significantly in terms of their contributions to prediction capabilities. Specifically, for the global SMIs data, their inclusion yields a notable enhancement in the prediction abilities of the forecasting modes across all three datasets: SSEC, SZEC, and GEI. This enhancement is particularly evident in the following five metrics: accuracy, recall, precision, \textit{F}-measure, and AUC. For the industry index data, their integration leads to overall improvements in prediction performance within the SZEC and GEI datasets. But integrating them within the SSEC dataset play a negative role on the prediction performance. For the news data, we find that the improvement degree is not obvious within the SSEC, SZEC, and GEI datasets by incorporating them in forecasting models. The second observation pertains to the comparative prediction performance of different models. The results in Table \ref{table:prediction results of different data sources} shows that SVM, RF, LSTM, and 2D-CNN models exhibit superior performance across SSEC, SZEC, and GEI datasets. Specifically, within the SSEC dataset, the SVM model achieves the highest frequency of optimal results, followed by the RF and LSTM model. In the SZEC dataset, the 2D-CNN model stands out as the most frequently superior method, followed by the SVM and CNN models. Similarly, within the GEI dataset, the LSTM model achieves the highest frequency of superior results, with the 2D-CNN and RF models trailing behind. 

\subsection*{C.3. The prediction results comparison of different prediction models}

\begin{table}[!ht]
    \centering
    \caption{Hyper-parameters settings of the TDSE-GA model}
    \label{table:hyper-parameters}
    \tiny
    \begin{tabularx}{\textwidth}
    {
        l
        l
        >{\raggedright\arraybackslash}X
        l
    }
    \hline
        Type & Model & Parameters & Value  \\ \hline
        Optimization model & GA & Population size & 50  \\ 
        ~ & ~ & Maximal generation number & 20  \\ 
        ~ & ~ & Maximal stall generation number & 5  \\ 
        ~ & ~ & Selection rate & 0.3  \\ 
        ~ & ~ & Crossover rate & 0.8  \\ 
    \hline
    Feature representation
     & MBCNN & Optimizer & Adam \\ 
        ~ & ~ & Activation function of convolutional layer and fully connected layer & RELU  \\ 
        ~ & ~ & Activation function of output layer & Softmax  \\ 
    
    ~ & SC-MBCNN & Method of determining the number of clusters & Elbow method  \\ 
    ~ & ~ & Optimizer & Adam  \\ 
    ~ & ~ & Activation function of convolutional layer and fully connected layer & RELU  \\ 
    ~ & ~ & Activation function of output layer & Softmax  \\ 
    
    ~ & RNN-ER & Optimizer & Adam  \\ 
    ~ & ~ & Activation function of output layer & Softmax  \\ \hline

    Meta learner & LR & C value & \{0.1, 1, 10, 100, 1000\}  \\ 
    ~ & RBF-SVM & C value & \{0.1, 1, 10, 100, 1000\}  \\ 
    ~ & ~ & Gamma value & \{0.1, 0.5, 1, 1.5, 2, 2.5\}  \\ 
    ~ & Poly-SVM & C value & \{0.1, 1, 10, 100, 1000\}  \\ 
    ~ & ~ & Polynomial value & \{2, 3, …, 10\}  \\ 
    ~ & RF & Number of decision trees & \{5, 10, 15, …, 50\}  \\ 
    ~ & ET & Number of decision trees & \{5, 10, 15, …, 50\}  \\ 
    ~ & ANN & Neuron numbers of each layer & \{2, 3, …, 50\}  \\ 
    ~ & KNN & The number of neighbor samples & \{2, 3, …, 20\}  \\ 
    \hline
    
    \end{tabularx}
\end{table}

\subsection*{C.4. Comparison of different optimization algorithms}

In this study, we employ the GA as our optimization algorithm. To validate the effectiveness of our chosen algorithm, we compare the GA with other optimization algorithms, specifically the Particle Swarm Optimization (PSO) algorithm, the Grey Wolf Optimizer (GWO), the Firefly Algorithm (FA), and the Harmony Search (HS) algorithm. The criteria for this comparison include the predictive classification performance and the computational time of each model. The prediction results of these various optimization methods on the SSEC, SZEC, and GEI datasets are shown in Table \ref{table:results of optimization methods} where the numbers in boldface within each column represent the highest values for that specific SMI. Based on the results in Table \ref{table:results of optimization methods}, we find the following two points. (1) The GA algorithm outperforms other optimization methods, providing the highest results in terms of the accuracy (SSEC and SZEC), recall (SSEC and GEI), precision (SSEC and GEI), \textit{F}-measure (SSEC, SZEC, and GEI), and AUC (SSEC and SZEC). (2) The FA algorithm provides suboptimal results in terms of the accuracy, recall, precision, \textit{F}-measure metric on SSEC. Similarly, the PSO algorithm shows suboptimal results in accuracy, precision, and AUC value on SZEC. Additionally, the GWO algorithm provides suboptimal results for accuracy, recall, precision, \textit{F}-measure, and AUC value on the GEI dataset.

\begin{table}[!ht]
\centering
\caption{The prediction results of different optimization methods on the SSEC, SZEC, and GEI}
\label{table:results of optimization methods}
\renewcommand\arraystretch{1.2}  
\begin{tabular}{lllllll}
\hline
SMI & Model & Accuracy & Recall & Precision & \textit{F}-measure & AUC  \\ \hline
\multirow{5}{*}{SSEC} 

~ & TDSE-PSO & 0.6059 & 0.6343 & 0.5885 & 0.5925 & 0.6042  \\ 
~ & TDSE-GWO & 0.6029 & 0.6134 & 0.5867 & 0.5822 & 0.5956  \\ 
~ & TDSE-FA & 0.6307 & 0.7021 & 0.6136 & 0.6265 & 0.6191  \\ 
~ & TDSE-HS & 0.5883 & 0.6372 & 0.5744 & 0.5852 & 0.5802  \\ 
& TDSE-GA & \textbf{0.6451} & \textbf{0.7236} & \textbf{0.6442} & \textbf{0.6521} & \textbf{0.6453}  \\ 
\hline

\multirow{5}{*}{SZEC} 
~ & TDSE-PSO & 0.6075 & 0.7291 & \textbf{0.6460} & 0.6290 & 0.6148  \\ 
~ & TDSE-GWO & 0.5917 & \textbf{0.7851} & 0.6171 & 0.6474 & 0.5932  \\ 
~ & TDSE-FA & 0.6035 & 0.7167 & 0.6164 & 0.6337 & 0.6009  \\ 
~ & TDSE-HS & 0.5987 & 0.7403 & 0.6358 & 0.6342 & 0.6051  \\ 
& TDSE-GA & \textbf{0.6225} & 0.7708 & 0.6312 & \textbf{0.6624} & \textbf{0.6192}  \\ 
\hline

\multirow{5}{*}{GEI} 

~ & TDSE-PSO & 0.5902 & 0.6236 & 0.6349 & 0.6188 & 0.5823  \\ 
~ & TDSE-GWO & \textbf{0.6273} & 0.7037 & 0.6719 & 0.6526 & \textbf{0.6263}  \\ 
~ & TDSE-FA & 0.6123 & 0.6612 & 0.6606 & 0.6363 & 0.6070  \\ 
~ & TDSE-HS & 0.6072 & 0.6819 & 0.6638 & 0.6402 & 0.6037  \\ 
& TDSE-GA & 0.6258 & \textbf{0.7419} & \textbf{0.6823} & \textbf{0.6663} & 0.6188  \\ 
\hline
\end{tabular}
\end{table}

\begin{figure}
    \centering
    \includegraphics[width=1\linewidth]{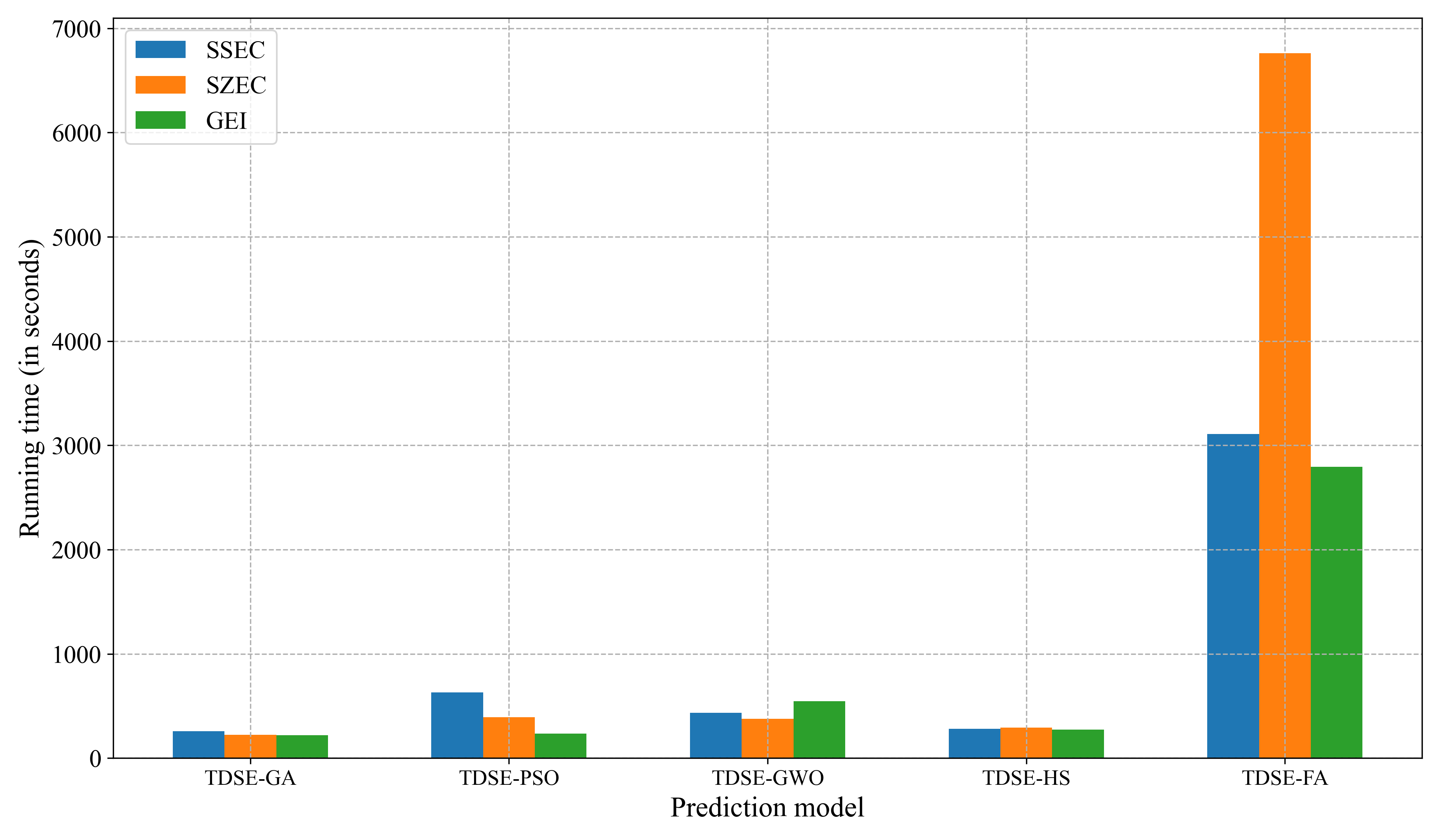}
    \caption{The running time of different optimization methods on SSEC, SZEC, and GEI}
    \label{fig:running times}
\end{figure}

In addition to comparing the prediction results, we also assess the computational efficiency by comparing the running time of different optimization algorithms. The running time of different optimization methods on the SSEC, SZEC, and GEI datasets is shown in Fig. \ref{fig:running times}. Based on Fig. \ref{fig:running times}, we first find that the GA algorithm exhibits shorter running times on the SSEC, SZEC, and GEI datasets compared to other optimization algorithms, with times of 257.15s, 221.47s, and 220.68s, respectively. Second, the FA algorithm consumes significantly longer running times on these datasets, with times of 3110.25s, 6760.76s, and 2792.34s, respectively. The longer running times for the FA algorithm can be attributed to the process of determining the optimal solution through pairwise comparisons between individuals in each generation. 

Based on the results presented in Table \ref{table:results of optimization methods} and Fig. \ref{fig:running times}, it is evident that the GA algorithm outperforms the PSO algorithm, the GWO algorithm, the HS algorithm, and the FA algorithm in both prediction performance and calculation efficiency. The GA algorithm consistently provides higher accuracy, recall, precision, \textit{F}-measure, and AUC values on the SSEC, SZEC, and GEI datasets compared to the other optimization methods. Additionally, the running time of the GA algorithm is significantly shorter, making it a more efficient choice for optimizing the second stage of the TDSE model. These experimental results emphasize the superiority of the GA algorithm as an optimization method in the proposed model.

\end{appendices}





\begin{spacing}{0.9}
    \footnotesize
    \bibliography{ref}
\end{spacing}


\end{document}